\documentclass[letterpaper,twocolumn,10pt]{article}
\usepackage{usenix}
\usepackage{amsmath,amssymb}
\usepackage{booktabs}
\usepackage{tabularx}
\usepackage{array}
\usepackage{url}
\usepackage{microtype}
\usepackage[table]{xcolor}
\usepackage{enumitem}
\usepackage{balance}
\definecolor{tablehead}{RGB}{224,236,248}
\definecolor{tablerow}{RGB}{247,250,253}
\definecolor{pathone}{RGB}{33,113,181}
\definecolor{pathtwo}{RGB}{0,109,44}
\definecolor{paththree}{RGB}{117,107,177}
\definecolor{pathfour}{RGB}{217,95,2}
\definecolor{pathfive}{RGB}{166,86,40}
\newcommand{\pone}{\textcolor{pathone}{\textbf{P1}}}
\newcommand{\ptwo}{\textcolor{pathtwo}{\textbf{P2}}}
\newcommand{\pthree}{\textcolor{paththree}{\textbf{P3}}}
\newcommand{\pfour}{\textcolor{pathfour}{\textbf{P4}}}
\newcommand{\pfive}{\textcolor{pathfive}{\textbf{P5}}}
\definecolor{modtab}{RGB}{94,60,153}
\definecolor{modimg}{RGB}{204,76,2}
\definecolor{modgraph}{RGB}{27,120,55}
\definecolor{modsensor}{RGB}{0,113,188}
\definecolor{scopelocal}{RGB}{63,63,63}
\definecolor{scopeglobal}{RGB}{110,59,125}
\definecolor{layertrain}{RGB}{0,109,44}
\definecolor{layerexplain}{RGB}{117,107,177}
\definecolor{layerpred}{RGB}{33,113,181}
\definecolor{layerapi}{RGB}{217,95,2}
\definecolor{layerreport}{RGB}{166,86,40}
\newcommand{\mtab}{\textcolor{modtab}{Tabular}}
\newcommand{\mimage}{\textcolor{modimg}{Image}}
\newcommand{\mgraph}{\textcolor{modgraph}{Graph}}
\newcommand{\msensor}{\textcolor{modsensor}{Sensor}}
\newcommand{\slocal}{\textcolor{scopelocal}{Local}}
\newcommand{\sglobal}{\textcolor{scopeglobal}{Global}}
\newcommand{\ltrain}{\textcolor{layertrain}{Training}}
\newcommand{\lexpl}{\textcolor{layerexplain}{Explainer}}
\newcommand{\lexplanation}{\textcolor{layerexplain}{Explanation}}
\newcommand{\lpred}{\textcolor{layerpred}{Prediction}}
\newcommand{\lapi}{\textcolor{layerapi}{API}}
\newcommand{\lreport}{\textcolor{layerreport}{Report}}
\newcolumntype{L}[1]{>{\raggedright\arraybackslash}p{#1}}
\newcolumntype{Y}{>{\raggedright\arraybackslash}X}

\title{SoK: Privacy Attacks on Machine Learning via Explainable AI}
\author{
{\rm Abdullah Caglar Oksuz}\\
Case Western Reserve University\\
abdullahcaglar.oksuz@case.edu
\and
{\rm Anisa Halimi}\\
IBM Research\\
anisa.halimi@ibm.com
\and
{\rm Erman Ayday}\\
Case Western Reserve University\\
erman.ayday@case.edu
}
\date{}

\begin{document}
\maketitle

\begin{abstract}
Machine learning explanations reveal model behavior beyond predictions, creating attack surfaces for model confidentiality and data privacy. We systematize 25 studies that exploit explanations for model extraction, membership inference, and model inversion, treating attribute inference as partial inversion. Existing work is often labeled only black- or white-box, obscuring substantial differences in what explanation signal reaches an adversary. We therefore separate model knowledge from explanation acquisition and identify five paths: target-released, attacker-derived, secondary disclosure, privileged access, and released global artifacts. Across these paths, explanations reduce extraction cost, expose membership signals through explanation statistics, recourse distance, and explanation-guided robustness, and support spatial or algebraic reconstruction of private inputs. We compare system and threat models, explanation signals, auxiliary knowledge, target models, modalities, query budgets, evaluation metrics, reported performance, and defenses. Our analysis shows that no explanation family is uniformly unsafe and no defense is uniformly effective. Risk depends on which signal is exposed, how it is acquired, which asset is targeted, and what the attacker already knows. We argue that explanation privacy should therefore be evaluated as an end-to-end disclosure problem, with defenses matched to the acquisition path and protected asset.
\end{abstract}

\section{Introduction}
\label{sec:intro}

Explainable artificial intelligence (XAI) releases information beyond the prediction itself. A feature-importance explanation identifies which dimensions mattered; a saliency map exposes spatial sensitivity; a counterfactual identifies a nearby point with a different outcome; a local surrogate approximates a decision region; and an influence-style explanation relates a prediction to training records. These artifacts serve legitimate goals such as debugging, recourse, auditing, and user understanding. They are useful precisely because they reveal structure beyond a label or confidence score. The same structure can become an attack surface.

A growing line of work has exploited explanations for three recurring goals. \emph{Model extraction} uses gradients, feature-importance explanations, counterfactuals, or local surrogates to reproduce model behavior with fewer or more informative queries~\cite{milli2019reconstruction,aivodji2020extraction,miura2021megex,oksuz2024autolycus}. \emph{Membership inference attacks (MIAs)} use explanation statistics, counterfactual distance, influence, or explanation-guided perturbations to decide whether a record was used for training~\cite{shokri2021privacy,pawelczyk2023recourse,liu2024tellmemore}. \emph{Model inversion} reconstructs all or part of a private record from explanation outputs, including sensitive-attribute inference as partial reconstruction~\cite{zhao2021exploiting,duddu2022attributes,luo2022feature,toma2024record}. These attacks affect different assets: model extraction primarily compromises model confidentiality, membership inference targets training-data privacy, and inversion may target either training records or private inputs submitted at inference time.

The gap is therefore not a missing catalog of explanation privacy risks. Recent surveys already organize attacks and privacy-preserving explanations~\cite{nguyen2025survey,allana2025scoping}, while SoKs on XAI in cybersecurity and adversarial settings cover broader offensive and defensive uses~\cite{nadeem2023sok,noppel2024sok}. The unresolved problem is \emph{comparability}: papers use similar labels for materially different security boundaries. ``Black-box'' may describe a label-only API, a prediction-plus-counterfactual service, an oracle returning the full input gradient, or a model with hidden parameters but a rich explanation tensor. These are not equivalent capabilities. An explanation may also come from the attacker's own query, a surrogate constructed by the attacker, another user's leaked report, or privileged model-owner access. Collapsing these settings obscures deployment realism and makes defense claims appear broader than the threat models they actually cover.

We retain the attack-family view expected by a privacy audience, namely model extraction, membership inference, and model inversion, but analyze every study through one system model. It separates (i) the protected asset and attack goal, (ii) model and training-data knowledge, (iii) prediction output, (iv) explanation acquisition, (v) auxiliary, shadow, and surrogate resources, and (vi) query budget and adaptivity. This makes it possible to compare modalities, model classes, metrics, and the incremental contribution of explanations without treating incompatible threat models as equivalent.

This analysis exposes four recurring security properties. First, \textbf{model knowledge and explanation disclosure are orthogonal}: hidden parameters do not prevent a service from exposing a gradient oracle with near-white-box learning value~\cite{milli2019reconstruction,miura2021megex}. Second, \textbf{the target need not release the explanation}: attacker-derived explanations can preserve attacks after an explanation endpoint is disabled~\cite{yan2023xamea,zhao2021exploiting}. Third, \textbf{explanation reports are sensitive derived data}: several inversion attacks target a report created for another user and later shared, intercepted, or compromised~\cite{luo2022feature,zhao2021exploiting,toma2024record}. Fourth, \textbf{defenses are path-specific}: differential privacy (DP), output perturbation, rate limiting, and access control protect different assets and disclosure channels.

\paragraph{Contributions.}
We make the following contributions:
\begin{itemize}[leftmargin=*,nosep]
    \item We systematize 25 core explanation-assisted attack studies across model extraction, membership inference, and model inversion, including attribute inference, while preserving the threat-model and metric differences needed to interpret their results.
    \item We introduce an \emph{explanation-acquisition} taxonomy comprising target-released, attacker-derived, secondary disclosure, privileged access, and released global artifact, separated from conventional model knowledge and prediction access.
    \item We compare the operational role of explanations, attack assumptions, modalities, metrics, and factors that determine leakage, rather than ranking attacks whose evaluation protocols are not comparable.
    \item We map defenses to the assets and acquisition paths they protect and derive reporting requirements for evaluating marginal leakage, low-FPR membership risk, extraction cost, and remaining disclosure channels.
\end{itemize}

\section{Scope and Methodology}
\label{sec:scope}

\subsection{Attack Scope}
We focus on attacks whose offensive procedure \emph{uses an explanation or interpretability artifact}. The three core families are:
\begin{description}[leftmargin=0pt,itemsep=2pt]
    \item[Model extraction.] The adversary learns a surrogate, parameters, or functionally equivalent model. Strictly, this targets model confidentiality rather than an individual's privacy. We retain it because the machine learning (ML) privacy literature commonly analyzes privacy and model confidentiality together~\cite{rigaki2023survey}, explanations are a direct extraction primitive, and an extracted surrogate can enable downstream membership or inversion attacks.
    \item[Membership inference.] The adversary decides whether a candidate record was used to train the target model.
    \item[Model inversion.] The adversary reconstructs private information represented by, or associated with, a record. We include sensitive-attribute inference as \emph{partial inversion}: the output is a hidden feature rather than a full record.
\end{description}

We exclude attacks that merely manipulate an explanation without a privacy or model-confidentiality objective (e.g., explanation evasion), generic adversarial examples, and gradient leakage from distributed training when the gradient is a training communication artifact rather than an explanation. Gradient-reconstruction attacks such as DLG and iDLG are relevant lineage for understanding gradient sensitivity, but they are outside the core corpus for this reason~\cite{zhu2019deep,zhao2020idlg,wainakh2021llg}. We also exclude prediction-only attacks except as baselines. Linkage and re-identification from counterfactuals are discussed as adjacent privacy objectives but are not forced into our three-family comparison~\cite{goethals2023linkage}.

\subsection{Corpus Construction and Cutoff}
Our objective is analytical systematization rather than a PRISMA-style review. Starting from the papers assembled for this manuscript, we cross-checked the closest surveys and SoKs~\cite{nguyen2025survey,allana2025scoping,nadeem2023sok,noppel2024sok} and used their references plus targeted freshness checks through \textbf{August 2, 2026} to identify missing work. A study enters the \emph{core attack corpus} if it evaluates one of our three attack families, the attack actually consumes an explanation signal, and the threat model is specific enough to classify the acquisition path. The final corpus contains \textbf{25 studies}: 13 on extraction, nine on membership inference, and four on inversion or attribute inference, with one study spanning extraction and membership inference. Recent preprints are retained only when they materially broaden the threat-model or modality landscape.

Unlike Allana et al.'s~\cite{allana2025scoping} broader 57-study scoping review and Nguyen et al.'s~\cite{nguyen2025survey} explanation-type taxonomy, our unit of analysis is the \emph{attack path} from explanation acquisition to a privacy or model-confidentiality objective. That narrower scope enables deeper comparison of operational assumptions and quantitative evidence.

\subsection{Coding Dimensions}
For each core study we record the attack family, protected asset, model knowledge, prediction output, explanation signal and acquisition path, auxiliary data, query budget, target model class, modality, success metric, representative result, and evaluated defenses. We do not force heterogeneous metrics into a synthetic ranking: 93\% extraction fidelity, 0.93 MIA AUC, and 0.90 image SSIM are different security properties. Instead, Sections~\ref{sec:mea}--\ref{sec:modia} compare results within each family and Section~\ref{sec:cross} identifies where evaluation choices block stronger conclusions.

\section{Background and Terminology}
\label{sec:background}

\subsection{Explanation Outputs}
Let a model $f_D$ be trained on dataset $D$. A conventional prediction service returns a prediction output $o(x)$, which may be a hard label or a confidence vector. An explanation mechanism additionally returns or makes available an artifact
\begin{equation}
    e(x) = \phi(f_D,x;R),
\end{equation}
where $R$ denotes any additional resources used by the explainer, such as a reference or background dataset, training examples, model gradients, or a counterfactual generator. For security analysis, $e(x)$ is not a single kind of output. Its content and the resources used to compute it determine what information may cross a trust boundary.

Feature-importance methods include model-agnostic local surrogates such as LIME~\cite{ribeiro2016lime}, game-theoretic attributions such as SHAP~\cite{lundberg2017shap}, and backpropagation-based methods for deep neural networks (DNNs), including Integrated Gradients~\cite{sundararajan2017axiomatic}, Grad-CAM~\cite{selvaraju2017gradcam}, and layer-wise relevance propagation (LRP)~\cite{bach2015lrp}. Counterfactual explanations (CFs) return a nearby input whose prediction differs and are commonly used for recourse~\cite{wachter2018counterfactual,mothilal2020dice}. Influence functions relate predictions or losses to training examples~\cite{koh2017influence}. These mechanisms expose different mathematical objects: importance vectors, local linear approximations, derivatives, boundary-adjacent examples, or training-record influence. Table~\ref{tab:xai-signals} summarizes the signals that appear in our corpus.

\begin{table*}[t!]
\centering
\caption{Explanation signals used in the core corpus. $\pone$--$\pfive$ denote the acquisition paths defined in Section~\ref{sec:threat}.}
\label{tab:xai-signals}
\scriptsize
\setlength{\tabcolsep}{2.7pt}
\renewcommand{\arraystretch}{1.08}
\rowcolors{2}{tablerow}{white}
\begin{tabularx}{\textwidth}{@{}L{1.45cm}L{1.52cm}L{2.0cm}L{2.25cm}L{1.25cm}Y@{}}
\rowcolor{tablehead}\toprule
\textbf{Signal} & \textbf{Typical scope} & \textbf{Model requirement} & \textbf{Security-relevant content} & \textbf{Acquisition path} & \textbf{Representative studies} \\
\midrule
Input gradients / saliency & \slocal & Differentiable model for direct computation & Local derivative, feature sensitivity, decision geometry & $\pone$, $\pfour$ & Milli et al.~\cite{milli2019reconstruction}; Miura et al.~\cite{miura2021megex}; Shokri et al.~\cite{shokri2021privacy} \\
IG / Grad-CAM / LRP & \slocal & Usually a differentiable neural network; method dependent & Spatial or feature relevance; class-specific structure & $\pone$, $\ptwo$, $\pthree$ & Zhao et al.~\cite{zhao2021exploiting}; Liu et al.~\cite{liu2024tellmemore}; Yan et al.~\cite{yan2022dtmea,yan2023xamea} \\
LIME & \slocal & Model-agnostic query access & Local feature weights and a local surrogate & $\pone$, $\ptwo$ & Oksuz et al.~\cite{oksuz2024autolycus}; Yan et al.~\cite{yan2023dmeae} \\
SHAP values & \slocal; \sglobal\newline aggregation & Model-specific or model-agnostic variants & Per-feature marginal contributions; correlations with private inputs & $\pone$, $\pthree$, $\ptwo$ & Oksuz et al.~\cite{oksuz2024autolycus}; Luo et al.~\cite{luo2022feature}; Toma and Kikuchi~\cite{toma2024record}; Duddu and Boutet~\cite{duddu2022attributes} \\
CF / recourse & \slocal & Prediction plus counterfactual-generation service & Boundary-adjacent instance, direction and magnitude of change & $\pone$ & A\"{\i}vodji et al.~\cite{aivodji2020extraction}; Wang et al.~\cite{wang2022dualcf}; Pawelczyk et al.~\cite{pawelczyk2023recourse}; Khouna et al.~\cite{khouna2025trees} \\
Influence & \slocal; training record & Usually model and training-process access & Contribution of training records to loss or prediction & $\pfour$ & Cohen et al.~\cite{cohen2024influence} \\
Global surrogate & \sglobal & Released or queryable surrogate & Compact approximation of global decision logic & $\pfive$ & Naretto et al.~\cite{naretto2022global} \\
Graph substructure & \slocal & GNN explainer dependent & Important nodes, edges or substructures & $\pone$ & Ma et al.~\cite{ma2025egsteal} \\
\bottomrule
\end{tabularx}
\end{table*}

\subsection{Marginal Leakage from Explanations}
The correct baseline for an explanation-assisted attack is the same adversary without the explanation signal. Conceptually, for an attack metric $M$ where higher is better,
\begin{equation}
 \Delta_{\mathrm{XAI}} = M(A\mid o,e,K,Q)-M(A\mid o,K,Q),
 \label{eq:delta}
\end{equation}
where $K$ is auxiliary knowledge and $Q$ is the query budget. For error metrics, the sign is reversed. Equation~\ref{eq:delta} is intentionally generic: the right family-specific metric may be fidelity at a fixed query budget, query savings at fixed fidelity, TPR at a fixed low FPR, or reconstruction error. The important requirement is to hold the remaining threat model fixed.

Many studies change more than the explanation channel itself, for example the optimizer, query strategy, or architecture. Liu et al.~\cite{liu2024tellmemore} show this explicitly: their stronger MIA depends on both XAI and a new multi-query procedure. By contrast, DTMEA~\cite{yan2022dtmea} gives a cleaner prediction-only versus prediction-plus-explanation comparison within one multitask design. Throughout this SoK we therefore distinguish \emph{attack performance with XAI} from evidence that isolates $\Delta_{\mathrm{XAI}}$.

\section{System and Threat Model}
\label{sec:threat}

\subsection{Model Knowledge and Interface Disclosure}
Traditional ML privacy work often summarizes attacker knowledge as black-box or white-box. For explanation-assisted attacks, this collapses two different dimensions. \emph{Model knowledge} asks whether parameters, architecture, gradients, or training data are known. \emph{Interface disclosure} asks what the deployed system returns. Milli et al.~\cite{milli2019reconstruction}, for example, study an unknown model through an oracle that returns function values and input gradients. The parameters remain hidden, yet the gradient can be a far stronger learning primitive than a hard label. MEGEX~\cite{miura2021megex} similarly treats the target model as externally queryable while using Vanilla Gradient explanations to recover the gradients needed for data-free distillation. Calling these simply ``black-box'' understates the information channel; calling them white-box incorrectly implies parameter access.

We therefore represent an attack setting as
\begin{equation}
 \mathcal{T}=\langle G,K_M,O,A_E,K_D,Q\rangle,
 \label{eq:tm}
\end{equation}
where $G$ is the attack goal and protected asset, $K_M$ is model knowledge, $O$ is the prediction output (label, score, etc.), $A_E$ is explanation acquisition, $K_D$ is auxiliary data and knowledge, and $Q$ captures query budget and adaptivity. This representation makes two attacks comparable only when the dimensions relevant to their goal are aligned.

\subsection{Explanation-Acquisition Paths}
Across the corpus, we identify five recurring ways in which an adversary obtains an explanation signal.

\paragraph{P1: Target-released.} The attacker sends a query and the deployed service returns the explanation for that query. Counterfactual extraction~\cite{aivodji2020extraction,wang2022dualcf,khouna2025trees}, explanation-guided MIA~\cite{shokri2021privacy,liu2024tellmemore}, and many gradient-oracle extraction studies assume this path. The core control point is the explanation endpoint and its output policy.

\paragraph{P2: Attacker-derived.} The target does not need to return the dangerous explanation. The attacker derives a local explanation from target predictions, trains a surrogate model or explainer, or reconstructs an explanation before using it in the privacy attack. XaMEA~\cite{yan2023xamea} includes attacks against targets that do not proactively provide explanations; Zhao et al.'s~\cite{zhao2021exploiting} surrogate-explanation pipeline can improve inversion even for a non-explainable target. This path survives a policy that merely disables explanation output.

\paragraph{P3: Secondary disclosure.} A legitimate explanation is generated for another data subject and subsequently shared, leaked, intercepted, or stolen. Luo et al.'s~\cite{luo2022feature} Shapley feature-inference framework first lets the attacker learn an inverse mapping with its own queries, then assumes access to target customers' explanation reports. Zhao et al.~\cite{zhao2021exploiting} model prediction and explanation tuples obtained from breached storage, interception, or social sharing; Toma and Kikuchi~\cite{toma2024record} likewise assume access to another user's Shapley values. Here the explanation report is itself sensitive derived data, and API rate limiting is not the primary defense.

\paragraph{P4: Privileged access.} The attack uses quantities normally available during training, debugging, or model-owner analysis rather than through a prediction service. Cohen et al.'s~\cite{cohen2024influence} self-influence MIA belongs here. Gradient leakage from distributed training also belongs to this broader privileged lineage, although it is outside our core corpus because the gradient is not an explanation artifact.

\paragraph{P5: Released global artifact.} A global surrogate or interpretable explanation model is released as an artifact. Naretto et al.~\cite{naretto2022global} evaluate membership exposure of global interpretable surrogates. The attack surface persists after release and is not governed by per-query access controls.

\subsection{Caller, Data Subject, and Adversary}
A separate ambiguity concerns identity. In P1 attacks, the adversary often queries its own record; in P3, it calibrates on its own queries but reconstructs a \emph{different person's} record from that person's report. Threat models should therefore name the \emph{caller}, the \emph{data subject}, and the \emph{adversary}. These roles may coincide, but assuming so can hide the true trust boundary. Stating only ``black-box'' is therefore incomplete: the model should also specify the prediction output, explanation artifact, acquisition path, auxiliary knowledge, and whose record is being explained.

\section{Model Extraction}
\label{sec:mea}

Model extraction is the largest attack family in the core corpus and exhibits the broadest range of explanation roles. The common objective is to learn a surrogate $\hat f$ that agrees with the target $f$ on a query distribution or to recover a functionally equivalent model. Explanations help because they expose more than a class label: a derivative can reveal local parameters, a counterfactual can locate a boundary, a feature-importance vector can guide which features to perturb, and an explanation-alignment loss can supervise the surrogate's decision logic. The literature evaluates these goals using target--surrogate fidelity, surrogate task performance, query count, and, in a smaller number of studies, similarity between target and surrogate explanations.

\begin{table*}[t!]
\centering
\caption{Model extraction attacks using explanations. Results follow each study's threat model and are not a single leaderboard.}
\label{tab:mea}
\scriptsize
\setlength{\tabcolsep}{2.3pt}
\renewcommand{\arraystretch}{1.07}
\rowcolors{2}{tablerow}{white}
\begin{tabularx}{\textwidth}{@{}L{1.34cm}L{1.63cm}L{1.63cm}L{1.35cm}L{1.65cm}L{1.45cm}Y@{}}
\rowcolor{tablehead}\toprule
\textbf{Study} & \textbf{Explanation (path)} & \textbf{Mechanism} & \textbf{Modality} & \textbf{Target models} & \textbf{Evaluation} & \textbf{Representative finding} \\
\midrule
Milli et al.~\cite{milli2019reconstruction} & Gradients or saliency; \pone & Parameter recovery; gradient matching & \mimage & Linear; two-layer ReLU; convolutional network; VGG11; ResNet-18 & Accuracy; queries & MNIST convolutional network: 95\% accuracy with 10 gradient queries versus 1,000 label queries; theory gives $O(h\log h)$ gradient queries. \\
A\"{\i}vodji et al.~\cite{aivodji2020extraction} & Prediction + CF; \pone & Surrogate training on factual and counterfactual pairs & \mtab & Logistic regression; multilayer perceptron & Fidelity; accuracy; queries & Adult: about 93\% fidelity with 1,000 queries under marginal-distribution knowledge; CFs remain useful with unknown distribution. \\
Kuppa and Le-Khac~\cite{kuppa2021adversarial} & Prediction + CF; \pone & CF-assisted coverage and distillation & \mtab\newline (security) & Gradient-boosted trees; neural networks & Surrogate accuracy; queries & CF-assisted extraction outperforms prediction-only and distillation baselines in the evaluated security setting. \\
Wang et al.~\cite{wang2022dualcf} & Prediction + CF + second CF; \pone & Two-sided boundary sampling & \mtab & Hidden-parameter classifiers & Fidelity; queries & Two-sided CF information improves extraction efficiency over one-sided CF extraction. \\
Yan et al.~\cite{yan2022dtmea} & Prediction + four feature-importance methods; \pone & Joint prediction and explanation training & \mimage & DNNs & Accuracy; explanation MSE & Prediction+explanation gains reach 1.25, 1.53, 9.25, and 7.45 points on MNIST, Fashion-MNIST, CIFAR-10, and CIFAR-100. \\
Yan et al.~\cite{yan2023dmeae} & Prediction + Grad-CAM or LIME; \pone & Data-free generation with explanation loss & \mimage & ResNet-34; ResNet-18 & Surrogate accuracy & Outperforms DFME and DaST on CIFAR-10 and SVHN; simple explanation perturbation changes attack accuracy by at most 0.65 percentage points. \\
Yan et al.~\cite{yan2023xamea} & Prediction + feature-importance map; \pone/\ptwo & Prediction and explanation fusion & \mimage & DNNs & Accuracy; fidelity & Explanation-assisted variants yield the largest gains on CIFAR-10 and CIFAR-100; attacker-derived explanations preserve part of the advantage. \\
Miura et al.~\cite{miura2021megex} & Soft prediction + gradient explanation; \pone & Gradient-guided data-free generation & \mimage & ResNet-34; ResNet-18; LeNet-5 & Surrogate accuracy; queries & CIFAR-10: 81.13\% accuracy at 2.5M queries versus 20M for reproduced DFME; 91.61\% at 20M queries. \\
Oksuz et al.~\cite{oksuz2024autolycus} & Label + LIME or SHAP; \pone & Explanation-guided active traversal & \mtab & Interpretable classifiers & Similarity; accuracy; queries & High similarity in several low-complexity settings; 100 queries for Breast Cancer logistic regression; pixel-level image explanations are ineffective. \\
Ezzeddine et al.~\cite{ezzeddine2024kd} & Prediction + private or non-private CF; \pone & Knowledge distillation with CFs & \mtab & DNNs & Agreement; accuracy & Non-private CFs improve extraction; DP-generated CFs reduce agreement toward no-CF baselines. \\
Khouna et al.~\cite{khouna2025trees} & Prediction + locally optimal CF; \pone & Tree Reconstruction Attack & \mtab & Decision trees; random forests & Fidelity; queries & Exact tree recovery with orders-of-magnitude fewer queries than PathFinding in many settings. \\
Ma et al.~\cite{ma2025egsteal} & Prediction + GNN explanation; \pone & Alignment and guided augmentation & \mgraph & Five GNN architectures & AUC; fidelity; rank correlation & NCI109 with Graph-CAM: +11.59 percentage points of fidelity over the prediction-only extraction baseline; gains vary by architecture and explainer. \\
Otto et al.~\cite{otto2026linear} & Factual + exact or robust CF; \pone & Geometric parameter recovery & \mtab & Linear classifiers & Exact recovery; queries & Exact parameter recovery from a small number of targeted queries under exact-oracle assumptions. \\
\bottomrule
\end{tabularx}
\end{table*}

\subsection{Gradient Explanations}
Milli et al.~\cite{milli2019reconstruction} provide the clearest demonstration that the \emph{information content} of an explanation matters more than the generic black-box label. For a linear model, one input-gradient query identifies the weight vector. For a two-layer ReLU network with $h$ hidden units under their stated independence assumptions, the model can be recovered with $O(h\log h)$ input-gradient queries, compared with an $\Omega(dh)$ information requirement for label queries in their analysis. Empirically, a MNIST convolutional model reaches 95\% accuracy after 10 gradient queries compared with 1,000 label queries; VGG11 and ResNet-18 still show roughly an order-of-magnitude reduction to reach a lower accuracy target. This is not conventional white-box access because the model parameters can remain secret, but the explanation oracle exposes a high-bandwidth learning primitive.

MEGEX~\cite{miura2021megex} pushes the same insight into data-free extraction. Data-free model extraction estimates gradients of the target indirectly through repeated prediction queries; MEGEX consumes Vanilla Gradient explanations directly and proves equivalence to the corresponding white-box data-free knowledge-distillation update under its formulation. On CIFAR-10, the reproduced DFME baseline reaches 81.13\% surrogate accuracy at 20M queries, while MEGEX reaches the same accuracy at 2.5M and 91.61\% at 20M. The absolute budgets are large, so this is not evidence that every gradient-explanation API is cheaply exploitable. It is evidence that returning raw or nearly raw derivatives can collapse a major source of query cost.

Processed gradients complicate any monotonic ``more interpretable means more leakage'' claim. MEGEX~\cite{miura2021megex} finds SmoothGrad and Integrated Gradients useful on Fashion-MNIST and SVHN but not equally effective on CIFAR-10. The key variable is not ``gradient-based XAI'' alone, but how faithfully the released explanation preserves the optimization signal the attacker needs.

\subsection{Counterfactual Explanations}
Counterfactual explanations are operationally attractive because they are designed to reveal a nearby point that changes the decision. This directly supplies boundary information. A\"{\i}vodji et al.~\cite{aivodji2020extraction} show that a surrogate can reach about 93\% fidelity on Adult with 1,000 queries even when the adversary knows only feature marginals; without knowledge of the data distribution, their CF-based attack can outperform a traditional extraction baseline using eight times more labels and full distribution knowledge. Providing multiple and diverse counterfactuals further improves extraction, illustrating a direct tension between explanation diversity for users and boundary coverage for attackers.

DualCF~\cite{wang2022dualcf} queries a counterfactual and then a counterfactual of that counterfactual, obtaining information from both sides of the boundary. Kuppa and Le-Khac~\cite{kuppa2021adversarial} use CFs to expand attack data across classes and combine them with distillation in cybersecurity models. More recent work moves from approximate to exact extraction. Khouna et al.~\cite{khouna2025trees} use their Tree Reconstruction Attack to recover decision trees and random forests from locally optimal counterfactuals, reaching functional equivalence with substantially fewer queries than PathFinding in many settings. Otto et al.~\cite{otto2026linear} analyze exact and robust counterfactual queries for linear classifiers and derive parameter-recovery results under explicit oracle assumptions.

This progression matters because later work shows that, for structured model classes, a CF oracle can be sufficient for exact functional recovery. Defenses based only on detecting high query volume therefore become less compelling as the required budget falls.

\subsection{Feature-Importance-Guided Extraction}
Not all extraction attacks treat the explanation as a direct parameter oracle. AUTOLYCUS~\cite{oksuz2024autolycus} uses LIME and SHAP to decide \emph{where to query}: explanation-identified important features and local boundaries guide active traversal, after which the attacker retrains a surrogate. This role is closer to active-learning guidance than gradient inversion. Across six tabular and mixed-type datasets and several interpretable model classes, the attack obtains high similarity with comparatively small budgets in favorable settings; for Breast Cancer logistic regression it reports high similarity with 100 queries. A direct query ratio against equation solving cannot be established for that dataset because the baseline does not report a corresponding query count. Its limitations are equally informative: pixel-wise LIME and SHAP are not sufficiently informative for the evaluated image setting, and complex trees require larger and more diverse auxiliary coverage. This is evidence that modality and semantic granularity of an explanation can dominate the nominal XAI method.

A series of attacks uses explanations as an auxiliary supervised target. DTMEA~\cite{yan2022dtmea} jointly learns predictions and explanations and reports larger gains on CIFAR-10 and CIFAR-100 than on MNIST and Fashion-MNIST. DMEAE~\cite{yan2023dmeae} uses an explanation loss in a data-free setting; XaMEA~\cite{yan2023xamea} fuses image explanations and predictions and also studies attacker-derived variants; and EGSteal~\cite{ma2025egsteal} transfers this idea to graphs. These attacks target not only output fidelity but aspects of \emph{decision logic}, so evaluations should state whether success means label agreement, parameter recovery, or explanation alignment.

\subsection{Evaluation of Model Extraction}
The dominant metrics are test accuracy of the surrogate, fidelity or agreement with the target, and query count. These capture different objectives. High test accuracy can coexist with low agreement when both models generalize differently; high fidelity on the natural test distribution can miss unobserved branches of a tree. Khouna et al.~\cite{khouna2025trees} therefore evaluate fidelity on uniformly sampled input-space points in addition to test-distribution evaluation. This is preferable when claiming functional recovery.

Query count is equally essential because the principal benefit of explanations is often efficiency. Reporting only the final surrogate accuracy can hide whether XAI saves 10 queries or millions. For learning-based extraction, we recommend an \emph{anytime fidelity curve} versus target queries and a fixed-fidelity query comparison. When one explanation request internally performs many model evaluations (e.g., model-agnostic SHAP or LIME), studies should also state whether the cost metric counts external requests or total target-model evaluations. Without that accounting, ``one explanation query'' can represent very different server-side work and information release.

Cross-paper query-efficiency claims also depend on baseline strength. Milli et al.~\cite{milli2019reconstruction} compare labels and gradients as different learning primitives, while MEGEX~\cite{miura2021megex} reproduces a data-free extraction baseline under closely matched architectures and training choices. A\"{\i}vodji et al.~\cite{aivodji2020extraction} explicitly vary architecture and distribution knowledge, showing that counterfactual information remains useful as those assumptions are weakened. Other comparisons change several factors at once, including prediction precision, auxiliary data, surrogate architecture, and query strategy. We therefore do not rank extraction methods across papers. A defensible explanation-versus-prediction comparison should hold these factors fixed and vary only the additional explanation channel.

The extraction literature supports a spectrum rather than a single claim: raw gradients can act as near-parameter oracles; counterfactuals expose boundary geometry; feature importance guides adaptive querying; and explanation alignment supplies an extra training target. The relevant security question is which learning primitive the interface reveals and at what query cost.

\section{Membership Inference}
\label{sec:mia}

Membership inference asks whether a record $z$ was included in the target training set. Explanations create membership signals in three main ways: (i) the explanation vector itself has different statistics for members and non-members; (ii) an explanation reveals distance or geometry related to the decision boundary; or (iii) it guides perturbations that expose a difference between members and non-members in robustness. These mechanisms are related to generalization and memorization, but they are not interchangeable. We retain each study's original metrics, while emphasizing true-positive rate (TPR) at a fixed false-positive rate (FPR) when available because it captures high-confidence leakage that average accuracy or area under the ROC curve (AUC) can obscure.

\begin{table*}[t!]
\centering
\caption{Membership inference attacks using explanations. Low-FPR results are highlighted when available.}
\label{tab:mia}
\scriptsize
\setlength{\tabcolsep}{2.3pt}
\renewcommand{\arraystretch}{1.07}
\rowcolors{2}{tablerow}{white}
\begin{tabularx}{\textwidth}{@{}L{1.4cm}L{1.65cm}L{1.8cm}L{1.3cm}L{1.58cm}L{1.42cm}Y@{}}
\rowcolor{tablehead}\toprule
\textbf{Study} & \textbf{Explanation (path)} & \textbf{Membership signal} & \textbf{Modality} & \textbf{Access} & \textbf{Evaluation} & \textbf{Representative finding} \\
\midrule
Shokri et al.~\cite{shokri2021privacy} & Gradients; Integrated Gradients; LRP; SmoothGrad; \pone & Explanation variance or full vector & \mtab; \mimage & Prediction + local explanation & Accuracy & Gradient explanations leak membership; full-vector attacks are slightly stronger than variance-only attacks, with leakage shaped by overfitting and dimensionality. \\
Kuppa and Le-Khac~\cite{kuppa2021adversarial} & CF; \pone & Proximity to factual and CF neighborhoods & \mtab\newline (security) & Prediction + CF API & Accuracy; queries & Demonstrates CF-assisted membership inference; its evaluation predates current low-FPR MIA practice. \\
Naretto et al.~\cite{naretto2022global} & Global surrogate; \pfive & Surrogate behavior & \mtab & Released global explainer & Accuracy; AUC & Global surrogate explanations can leak more membership information than the original classifier; leakage tracks fidelity and overfitting. \\
Pawelczyk et al.~\cite{pawelczyk2023recourse} & CF or recourse; \pone & CF distance; likelihood ratio & \mtab & One recourse query per record & AUC; balanced accuracy; low-FPR TPR & HELOC+CCHVAE: AUC 0.679 and TPR 5.13\% at 1\% FPR; several Adult and Diabetes settings remain near random. \\
Cohen et al.~\cite{cohen2024influence} & Self-influence; \pfour & Self-influence score & \mimage & Model and training access & Accuracy; AUC & Demonstrates membership leakage from self-influence under a substantially stronger, non-public access model. \\
Liu et al.~\cite{liu2024tellmemore} & Seven feature-importance methods; \pone/\ptwo & Explanation-guided robustness trajectory & \mimage & Prediction + explanation; shadow variants & AUC; balanced accuracy; TPR at 0.1\% FPR & Representative AUC 0.915--0.931 and TPR 2.2--3.7\% at 0.1\% FPR; ablations show both XAI and the multi-query procedure contribute. \\
Ma et al.~\cite{ma2024labelonly} & Feature-importance signal; \ptwo & Neighborhood consistency and recovered confidence & \mimage & Label-only; 500-query cap & Accuracy; low-FPR TPR; queries & Feature importance guides sampling and supports high-precision attacks on selected forgettable records, trading coverage for precision. \\
Ezzeddine et al.~\cite{ezzeddine2026leak} & CF; \pone & CF-augmented attack features & \mtab; \msensor & Prediction + CF; DP and active-learning variants & Accuracy; precision; recall & EEG baseline: accuracy and recall rise from 58\% without CF to 77\% with CF; DP reduces leakage with a utility cost. \\
Benhmida et al.~\cite{benhmida2026deepleak} & 15 methods across four explanation families; \pone & Learned explanation-based signal & \mimage & Prediction + explanation & Membership leakage; explanation utility & Reports up to 74.9\% more leakage than prior explanation-MIA baselines; mitigations reduce leakage by 46.5--95\% with at most 3.3\% average utility loss. \\
\bottomrule
\end{tabularx}
\end{table*}

\subsection{Explanation Statistics}
Shokri et al.~\cite{shokri2021privacy} provide the foundational analysis of membership leakage through feature-based explanations. For backpropagation-based explanations, variance (or equivalently related norms in their experiments) can separate members from non-members because explanation magnitude or variance relate to a point's position relative to the learned boundary. A learned attack on the entire explanation vector is slightly stronger than a variance-only threshold. Integrated Gradients and LRP also leak membership, although generally less than raw gradient explanations in their evaluated settings. Importantly, leakage varies with input dimensionality, number of classes, and overfitting; ``gradient explanation = vulnerable'' is not a constant attack rate.

Global explanations introduce a different channel. Naretto et al.~\cite{naretto2022global} show that an interpretable global surrogate can be more vulnerable to MIA than the original black-box model. The risk here persists after the surrogate is released and is governed by how the global explainer approximates or overfits the target, not by repeated local explanation queries.

Cohen et al.~\cite{cohen2024influence} use self-influence to identify membership. The result is relevant to explanation leakage but has a substantially stronger access model: self-influence is computed with privileged model and training information. We therefore classify it separately from public explanation APIs. Combining it with external attacks under a single ``white-box versus black-box XAI'' ranking would obscure deployability.

\subsection{Counterfactual Distance}
Algorithmic recourse creates a natural geometric statistic: the distance from a point to a counterfactual. Pawelczyk et al.~\cite{pawelczyk2023recourse} formalize counterfactual-distance (CFD) attacks and a likelihood-ratio version using this distance. Their results are notable for heterogeneity. On HELOC with CCHVAE recourse, the LRT reaches AUC 0.679 and TPR 5.13\% at 1\% FPR; several Adult and Diabetes configurations remain close to random. Thus the existence of a CF API does not imply uniformly strong MIA. Leakage depends on the model, data geometry, and recourse algorithm.

Kuppa and Le-Khac~\cite{kuppa2021adversarial} also use counterfactual neighborhoods in a black-box membership attack. Ezzeddine et al.~\cite{ezzeddine2026leak} later show a sizeable empirical increase when counterfactual information is added to an MIA on EEG: accuracy and recall rise from 58\% without CFs to 77\% with CFs in the reported baseline setting. Because the latter is a preprint and uses a different evaluation protocol, it should be read as evidence that CF information can amplify leakage, not as a direct ranking over Pawelczyk's attacks.

\subsection{Explanation-Guided Robustness}
Liu et al.~\cite{liu2024tellmemore} shift the MIA signal from the explanation vector to \emph{model robustness under explanation-guided perturbation}. The attacker ranks pixels using SmoothGrad, VarGrad, Integrated Gradients, Grad-CAM, Grad-CAM++, LIME, or SHAP, perturbs most- or least-relevant regions, and classifies the resulting confidence-degradation trajectory. Across the seven explanation methods in a representative CIFAR-100 setting, reported AUCs are roughly 0.915--0.931 and TPR at 0.1\% FPR is 2.2--3.7\%. These numbers are strong, but the paper's ablations are crucial: an explanation-free segmentation and perturbation strategy also improves substantially over older baselines. The result therefore demonstrates a strong \emph{explanation-assisted attack pipeline}; its raw improvement should not be attributed entirely to XAI.

Ma et al.~\cite{ma2024labelonly} pursue a related idea under stricter label-only output. Feature importance guides high-dimensional neighborhood sampling, from which the attacker derives an attributed-neighborhood consistency signal and approximates confidence. The attack explicitly selects ``forgettable'' or vulnerable records to obtain high precision at low FPR while sacrificing coverage. This is a meaningful privacy threat because an attacker may only need to identify some users with high confidence, but the coverage trade-off must accompany the headline TPR.

DeepLeak~\cite{benhmida2026deepleak} broadens the explanation-method coverage and directly studies leakage drivers and mitigations across 15 XAI techniques. For this SoK, the important result is broader than the reported MIA improvement: explanation properties such as sensitivity and sparsity predict leakage more directly than coarse method labels, and the study evaluates output-level mitigations across multiple explanation families.

\subsection{Low-FPR Evaluation}
Average accuracy and AUC remain useful for comparison, but privacy attacks are often operationally important in the tail. If an attack identifies 1\% of training records with very high confidence and guesses randomly on the rest, its average accuracy can appear benign. Following modern MIA practice, including Pawelczyk et al.~\cite{pawelczyk2023recourse}, Liu et al.~\cite{liu2024tellmemore}, and Ma et al.~\cite{ma2024labelonly}, studies should report TPR at fixed low FPR (at least 1\% and, where sample size permits, 0.1\% or lower), alongside AUC and balanced accuracy. The number of positive and negative evaluation examples must be large enough to support the claimed operating point.

A second reporting issue is per-record versus population-average leakage. Explanation MIA repeatedly finds that outliers, difficult examples, or forgettable records leak more. A single average can therefore hide highly vulnerable subpopulations. Future evaluations should report distributions or stratified leakage, not only one scalar.

Target selection must be reported with the same care. Pawelczyk et al.~\cite{pawelczyk2023recourse} show recourse configurations that remain close to random even when others leak substantially, whereas Ma et al.~\cite{ma2024labelonly} deliberately focus on records estimated to be especially vulnerable. These results answer different questions: population-wide risk versus the ability to identify a high-risk subset. When an attack filters candidate records before inference, the paper should report coverage, the selection rule, and whether the information needed for selection is available under the stated threat model. Otherwise, a high TPR can conceal that the attack applies to only a narrow subset or relies on additional knowledge.

Explanation-assisted membership inference spans several mechanisms. Direct explanation statistics, boundary or recourse distance, and explanation-guided robustness each leak under different conditions. Strong evaluation therefore requires low-FPR metrics, coverage, and an ablation that separates the explanation's contribution from the attack algorithm's contribution.

\section{Model Inversion}
\label{sec:modia}

Model inversion presents the most direct personal-data risk in the corpus: explanations are used to reconstruct a private record or a hidden attribute. A critical distinction is \emph{whose data are reconstructed}. Classic model inversion often aims at information memorized in the training set~\cite{fredrikson2015model}; several explanation attacks instead reconstruct the \emph{query input} represented by an explanation report. The latter risk exists even if the model never memorized that record. Accordingly, the literature uses different evaluation measures: mean absolute error (MAE) for tabular reconstruction, mean-squared error (MSE), structural similarity (SSIM), and peak signal-to-noise ratio (PSNR) for images, and classification accuracy for hidden attributes.

\begin{table*}[t!]
\centering
\caption{Comparison of explanation-assisted model inversion studies, including attribute inference as partial inversion. We distinguish reconstruction of training records from reconstruction of query records.}
\label{tab:modia}
\scriptsize
\setlength{\tabcolsep}{3.2pt}
\renewcommand{\arraystretch}{1.10}
\rowcolors{2}{tablerow}{white}
\begin{tabularx}{\textwidth}{@{}L{1.45cm}L{1.85cm}L{2.20cm}L{1.55cm}L{1.7cm}L{1.6cm}Y@{}}
\rowcolor{tablehead}\toprule
\textbf{Study} & \textbf{Explanation (path)} & \textbf{Mechanism} & \textbf{Private target} & \textbf{Target models} & \textbf{Evaluation} & \textbf{Representative finding} \\
\midrule
Zhao et al.~\cite{zhao2021exploiting} & Gradient, CAM, and LRP; \pone/\ptwo/\pthree & Spatial explanation as inversion input & \mimage{} records & DNN image classifiers & MSE; SSIM; PSNR; attack accuracy & Spatial explanations improve over prediction-only inversion; richer class-specific CAM sets increase reconstruction risk. \\
Duddu and Boutet~\cite{duddu2022attributes} & Feature-importance explanations; \pone & Classifier maps explanation to hidden attribute & \mtab{} records & Four classifier families & Attribute-inference accuracy & Explanations outperform prediction-only inference; leakage remains when the sensitive feature is censored via proxy correlations. \\
Luo et al.~\cite{luo2022feature} & Shapley report; \pone~(calibration), \pthree~(target) & Learned inverse from local linearity & \mtab{} query records & Neural network; support vector machine; random forest; gradient-boosted tree & MAE; success rate & Demonstrated on Google, Microsoft, and IBM services; important features and neural networks are generally more vulnerable. \\
Toma and Kikuchi~\cite{toma2024record} & Shapley values; \pthree & Inverse optimization; exact linear case & \mtab{} query records & Linear and nonlinear models & Error; exact recovery & Exact reconstruction for linear regression with exact Shapley values; empirical risk varies by model and optimizer. \\
\bottomrule
\end{tabularx}
\end{table*}

\subsection{Spatial Explanations}
Zhao et al.~\cite{zhao2021exploiting} show that image explanations can serve as a spatial prior for instance-level inversion. Their multimodal transposed-CNN and U-Net architectures consume a prediction and a saliency or CAM-style map. U-Net bypass connections preserve spatial information that a flattened explanation loses, increasing reconstruction quality under MSE, SSIM, PSNR, embedding similarity, and downstream attack accuracy. Richer sets of class-specific CAM explanations further increase risk. The security implication is that a saliency map reveals more than feature relevance; its two-dimensional structure can encode enough information about the input to improve image recovery.

The threat model is also notable. Zhao et al.~\cite{zhao2021exploiting} consider prediction and explanation tuples obtained through breached storage, interception, or social sharing, and an independent auxiliary dataset plus black-box API access. This is P3 secondary disclosure, not merely a malicious caller eliciting another person's explanation. The paper also reconstructs surrogate explanations when the target itself does not expose them, creating a P2 variant. A defense that only authenticates the explanation API addresses the first stage of P1 but not these alternative paths.

\subsection{Shapley-Value Inversion}
Luo et al.~\cite{luo2022feature} study Shapley-based feature inference against tabular ML services. The attacker first sends its own inputs to learn the relationship between features and Shapley explanations; it then obtains a target user's explanation report and reconstructs that user's private input features. The paper evaluates Google Cloud, Microsoft Azure, and IBM AIX360-era services and reports MAE and success rate across neural-network, support-vector-machine, random-forest, and gradient-boosted-tree targets. Neural networks are generally most vulnerable in their experiments, and features with greater model importance are reconstructed more reliably. The attack does not rely on the target record being in the training set. Consequently, training-time anti-memorization defenses do not address the central leakage channel.

Toma and Kikuchi~\cite{toma2024record} sharpen the same issue theoretically. With exact Shapley values and a linear-regression target, the private input can be reconstructed without error under their model; experiments on Adult, Bank Marketing, and Credit Card Client examine optimizer and model combinations and approximate settings. The result shows why it is unsafe to speak about ``SHAP privacy'' without specifying the model and implementation: an exact algebraic relationship in one model family may become an approximate statistical inversion problem in another.

\subsection{Attribute Inference as Partial Inversion}
Duddu and Boutet~\cite{duddu2022attributes} infer sensitive attributes such as race or sex from explanations. Their two threat models are especially useful for privacy reasoning. In one, the sensitive attribute appears in model input and training data; in the other, it is censored entirely. Even in the censored setting, proxy features can make explanations distinguishable by the hidden sensitive attribute. Across four datasets and four model families, explanations alone outperform prediction-only attribute inference in the studied settings; concatenating predictions with explanations does not consistently help and can hurt. Thus the explanation can be the primary privacy channel rather than merely an amplifier of confidence-based inference.

Calling attribute inference a subcategory of inversion is useful operationally if we preserve the output distinction: full inversion reconstructs a record or high-dimensional representation, while attribute inference reconstructs selected hidden components. The defense target is the same class of asset, namely private information about an individual input, but the evaluation metric differs. Attribute accuracy must be compared against class-prior and proxy-feature baselines; image inversion needs similarity and semantic metrics rather than raw classification accuracy.

\subsection{Evaluation of Reconstruction Risk}
Pixel MSE, PSNR, and SSIM, feature MAE, success rate, and attribute accuracy measure different notions of recovery. A visually imperfect face may still preserve identity, and a low average tabular MAE may hide exact reconstruction of a particularly sensitive feature. Conversely, high SSIM may largely reflect background structure irrelevant to the private property. We therefore recommend pairing reconstruction fidelity with a task-specific privacy metric: re-identification or sensitive-attribute accuracy for images, per-feature success with explicit sensitive-feature analysis for tabular data, and prior-normalized accuracy for attribute inference.

The attacker's prior is equally important. Luo et al.~\cite{luo2022feature} evaluate both an adversary with auxiliary data from the input distribution and a weaker adversary without that background knowledge; Duddu and Boutet~\cite{duddu2022attributes} compare explanations against prediction-based attribute inference. Reconstruction results should therefore be reported as improvement over an appropriate prior or prediction-only baseline, not only as absolute error. For Shapley-based attacks, the implementation also matters: Toma and Kikuchi's~\cite{toma2024record} exact result assumes exact Shapley values and a linear model, whereas commercial systems generally return approximations. A theoretical exact-recovery result and an empirical attack on approximate reports are complementary evidence, not interchangeable threat models.

Explanation-assisted inversion exposes two distinct privacy problems: leakage about training records and leakage about a live query record. Shapley and image-report attacks show that the latter can arise after a legitimate explanation is generated, making report handling and secondary disclosure part of the threat model.

\section{Cross-Attack Analysis}
\label{sec:cross}

\subsection{How Explanations Change the Attack Surface}
Across the corpus, explanations play six recurring operational roles: gradient oracle, boundary localization, query guidance, surrogate supervision, membership statistic, and inverse representation. This vocabulary is more informative than calling XAI a primary vector or an amplifier because it names the information the attacker actually uses.

The acquisition path then determines whether that information is exposed by the deployment. P1 relies on a target-released explanation interface; P2 survives if the attacker can derive the signal independently; P3 turns the explanation report itself into sensitive derived data; P4 requires privileged training or model access; and P5 persists after a global artifact has been released. We therefore avoid labeling SHAP, LIME, gradients, or counterfactuals as inherently ``end-user'' or ``owner-only.'' Exposure is a property of the deployment, authorization boundary, and recipient. These paths also explain defense failures: feature-level noise may break query guidance while leaving a gradient oracle useful; blurring a counterfactual may reduce boundary information but also degrade recourse; and sanitizing a released explanation does not stop P2 attacks that derive explanations from predictions. A realistic comparison must therefore state \emph{which} assumption changes: model knowledge, explanation disclosure, auxiliary data, caller identity, or query capability.

\subsection{Factors Affecting Leakage}
The corpus does not support a universal ranking of explanation methods by privacy risk. Leakage depends first on how closely the released signal matches the attack primitive. Raw gradients can be unusually effective for extraction~\cite{milli2019reconstruction}, while processed gradients can remove enough information to reduce MEGEX~\cite{miura2021megex} performance on harder data. Counterfactual extraction benefits from boundary proximity and diversity~\cite{aivodji2020extraction,wang2022dualcf,khouna2025trees}, and inversion benefits from spatially structured explanations when the target is an image~\cite{zhao2021exploiting}. The security question is therefore not whether an explanation is ``faithful'' in the abstract, but whether it preserves the statistic required by a particular attack.

Model complexity, auxiliary knowledge, and modality are the other recurring factors. Query savings can shrink as targets become more complex~\cite{milli2019reconstruction}; complex trees and weakly informative feature-level explanations require broader coverage in AUTOLYCUS~\cite{oksuz2024autolycus}; exact Shapley values can permit algebraic inversion in linear models~\cite{toma2024record}; and EGSteal~\cite{ma2025egsteal} reports architecture- and explainer-dependent gains on graphs. Counterfactual extraction remains effective under reduced distribution knowledge~\cite{aivodji2020extraction}, membership leakage is shaped by overfitting and record heterogeneity~\cite{shokri2021privacy,pawelczyk2023recourse,ma2024labelonly}, and Shapley-based query-record inversion need not involve a training point at all~\cite{luo2022feature}. Evidence also remains concentrated in tabular and image settings, with much less mature coverage for audio, sequential data, and text or LLM interfaces.

\paragraph{Which explanation types dominate?} The corpus is strongly skewed toward \emph{local} explanations: counterfactual/recourse mechanisms appear in eight core studies, while SHAP/Shapley and raw gradients each appear in six. Model-agnostic feature importance and counterfactuals dominate tabular attacks, whereas gradient, CAM, and relevance maps dominate image attacks; graph explanations appear only in recent extraction work. Defense research mirrors this skew, concentrating on feature-importance perturbation, DP explanations, and counterfactual sanitization. Current evidence therefore characterizes local, record-specific explanation channels far better than global or graph explanations.

\subsection{Interpreting Reported Results}
Several studies find greater leakage when an explanation exposes more attack-relevant information: diverse counterfactuals improve extraction~\cite{aivodji2020extraction}, multiple class-specific CAMs improve inversion~\cite{zhao2021exploiting}, and raw gradients sharply reduce extraction query requirements~\cite{milli2019reconstruction}. These observations do not justify a monotonic rule that richer or more faithful explanations always leak more. Processed gradients can weaken MEGEX~\cite{miura2021megex} in some settings; perturbation-based explanations can leak less membership information than raw gradients in Shokri et al.'s~\cite{shokri2021privacy} experiments; and an explanation that is useful to a human may discard exactly the numerical structure an attack needs. A more defensible conclusion is conditional: additional explanation detail increases risk when it preserves information aligned with the attack's decision statistic or optimization objective.

A second limitation is comparability. Model-extraction papers report fidelity, task accuracy, and query counts; MIA papers increasingly emphasize TPR at low FPR; inversion papers use feature error, image similarity, or attribute accuracy. Even ``one query'' is not a consistent unit: a single request for a model-agnostic explanation may trigger many internal model evaluations, while one gradient response can return a high-dimensional vector. We therefore treat reported numbers as evidence within each paper's threat model rather than as a cross-paper leaderboard. A stronger evaluation should report the external request count, response dimensionality and precision, the number of explanation alternatives returned, and, where it can be measured, the internal target-evaluation cost.

Most importantly, attack performance with explanations is not automatically evidence of \emph{marginal leakage caused by explanations}. A new attack can improve because of a better optimizer, more adaptive queries, a stronger attack model, or additional auxiliary data. Liu et al.'s~\cite{liu2024tellmemore} explanation-free ablation demonstrates this directly: both the explanation and the multi-query perturbation procedure contribute to the final MIA. Results that isolate Eq.~\ref{eq:delta}, such as matched prediction-only versus prediction-plus-explanation variants, are therefore more informative for deciding whether an explanation interface should be exposed.

\subsection{Recommended Reporting Practice}
Based on these inconsistencies, future explanation-assisted attack evaluations should report:
\begin{enumerate}[leftmargin=*,nosep]
    \item The protected asset and exact attack output;
    \item Model knowledge separately from the prediction output;
    \item The explanation artifact and its P1--P5 acquisition path;
    \item Whether the caller, data subject, and adversary are the same party;
    \item Auxiliary or shadow data assumptions and distribution match;
    \item External requests and, where meaningful, internal target-evaluation cost;
    \item A prediction-only baseline with the remaining attack components held fixed;
    \item Family-appropriate metrics, including low-FPR TPR for MIA and fidelity-versus-query curves for extraction;
    \item Performance heterogeneity across records, model classes, and modalities;
    \item Evaluated defenses together with any disclosure channels that remain open.
\end{enumerate}

\section{Defenses}
\label{sec:defenses}

DP, explanation perturbation, query controls, and access control operate at different layers and protect different assets. We include differentially private stochastic gradient descent (DP-SGD) as a training-time defense, but distinguish its training-record guarantee from mechanisms that protect the explanation or the query record. We group defenses by the layer on which they act and ask which acquisition path they can actually close. Table~\ref{tab:defenses} summarizes the evidence.

\begin{table*}[t!]
\centering
\caption{Defenses relevant to explanation-assisted attacks. Evidence is classified as direct empirical evaluation, a formal privacy argument, or an unvalidated proposal for the corresponding attack family.}
\label{tab:defenses}
\scriptsize
\setlength{\tabcolsep}{2.7pt}
\renewcommand{\arraystretch}{1.09}
\rowcolors{2}{tablerow}{white}
\begin{tabularx}{\textwidth}{@{}L{1.95cm}L{1.52cm}L{1.65cm}L{1.95cm}L{1.85cm}Y@{}}
\rowcolor{tablehead}\toprule
\textbf{Defense} & \textbf{Layer} & \textbf{Primary asset} & \textbf{Evidence} & \textbf{Not automatically covered} & \textbf{Main observation} \\
\midrule
DP-SGD~\cite{dpsgd} & \ltrain & Training membership and records & Formal and empirical~\cite{patel2022dp,liu2024tellmemore,pawelczyk2023recourse} & Query-record privacy; model confidentiality & When explanations are post-processing of a DP model, the training-data guarantee is preserved; practical utility and explanation-quality loss can be substantial. \\
Regularization and dropout~\cite{dropout} & \ltrain & Empirical membership leakage & Indirect or general evidence & Formal privacy; query inversion; model extraction & Can reduce overfitting-driven leakage but provides no formal privacy guarantee. \\
DP local explanations~\cite{patel2022dp} & \lexpl & Reference or training data & Formal and empirical analysis & Model confidentiality; unprotected query records & Separates DP of model training from privacy of data used to construct local explanations. \\
DP counterfactuals~\cite{ezzeddine2024kd} & \lexpl & Model confidentiality & Model extraction: evaluated & Attacker-derived explanations; other XAI families & Reduces CF-assisted extraction toward no-CF baselines while retaining some recourse utility. \\
Noise on feature explanations~\cite{allana2025noise} & \lexplanation & Query-record attributes & Inversion; attribute inference: evaluated & Raw gradients; attacker-derived channels & Evaluates attack success, explanation faithfulness, and model utility jointly. \\
Clipping, masking, and noise~\cite{benhmida2026deepleak} & \lexplanation & Membership leakage & Membership inference: evaluated & Extraction; inversion unless evaluated & Reduces leakage by 46.5--95\% with at most 3.3\% average utility loss in DeepLeak. \\
FGSM-style explanation perturbation~\cite{yan2023dmeae} & \lexplanation & Model confidentiality & Model extraction: evaluated & Other attack families; adaptive recovery & Changes DMEAE attack accuracy by only 0.49--0.65 percentage points in the reported settings. \\
MemGuard~\cite{memguard} & \lpred & Prediction-based membership leakage & Membership inference: evaluated~\cite{liu2024tellmemore} & Explanation channel & Explanation-assisted attacks retain AUC 0.872--0.885 because feature-importance outputs remain unprotected. \\
Query budgets and auditing (PRADA)~\cite{prada} & \lapi & Repeated online extraction & Partial or indirect evidence~\cite{aivodji2020extraction,oksuz2024autolycus} & Low-query and Sybil attacks; secondary disclosure & Raises attack cost, but low-query CF extraction can resemble benign use and queries can be distributed across identities. \\
Access control and retention & \lreport & Query-record privacy & Proposed in XAI reviews~\cite{allana2025scoping} & Legitimate sharing; attacker-derived explanations & Directly addresses secondary disclosure of sensitive explanation reports. \\
Faithful\allowbreak Defense~\cite{zhong2025faithful} & \lexpl & Model confidentiality & Model-extraction-oriented design & Training and query-record privacy & Reduces boundary leakage while retaining faithful logical explanations. \\
XRand~\cite{nguyen2023xrand} & \lexplanation & Feature-importance disclosure & Evaluated for backdoor guidance & Extraction; MIA; inversion unvalidated & Useful local-DP-style randomization precedent, but published evaluation targets a different attack family. \\
\bottomrule
\end{tabularx}
\end{table*}

\subsection{Training-Time Defenses}
Differentially private training is the strongest general mechanism in our corpus for limiting information about individual training records. Patel et al.~\cite{patel2022dp} make the crucial post-processing argument explicit: if $f_D$ is differentially private with respect to training dataset $D$, then an explanation computed only from that DP output or model cannot worsen the formal training-data guarantee. Their work separately protects the private \emph{explanation or reference data} used to construct a local explanation, which may not be covered by DP model training.

This distinction prevents an important overclaim. DP-SGD can bound training-membership leakage, but it does not automatically protect a new user's private query from a Shapley inversion attack, because that query is not the protected training record. Nor is DP designed to preserve proprietary model confidentiality: a DP model can still be functionally extracted. Pawelczyk et al.~\cite{pawelczyk2023recourse} further note that DP training may move decision boundaries and make recourse less actionable; Liu et al.~\cite{liu2024tellmemore} find that practical DP-SGD configurations reduce MIA but can substantially damage both classification and explanation quality. Formal protection and usable explainability therefore need joint evaluation.

Regularization, dropout, augmentation, and early stopping are useful empirical defenses when MIA is driven by overfitting, but they are not privacy guarantees~\cite{dropout,rigaki2023survey}. They are even less directly matched to query-record inversion or model extraction. Their place in an XAI privacy paper should be as general baselines, not as substitutes for explanation-specific controls.

\subsection{Explanation-Layer Defenses}
Explanation-layer defenses perturb or redesign the artifact before release. Patel et al.'s~\cite{patel2022dp} DP local explanations provide formal protection for data used during local approximation. Allana and Dara~\cite{allana2025noise} add noise to feature-based explanations and evaluate attribute-inference risk, explanation faithfulness, and predictive utility jointly. DeepLeak~\cite{benhmida2026deepleak} evaluates clipping, masking, and calibrated noise over a broader set of explainers. These works point toward the correct optimization target: not ``make the explanation noisy,'' but remove the attack-relevant statistic while retaining the utility property required by the intended recipient.

The difficulty is attack specificity. DMEAE~\cite{yan2023dmeae} evaluates a simple FGSM-style perturbation of explanations and observes only 0.49 percentage-point change in SVHN attack accuracy and 0.65 points on CIFAR-10. Perturbation magnitude that visually changes an explanation may leave enough structure for extraction. Conversely, aggressive noise that defeats MIA can make an explanation useless. An explanation defense should therefore report both attack success for the corresponding family and a method-appropriate explanation-quality metric.

Counterfactual privacy adds another trade-off. Ezzeddine et al.~\cite{ezzeddine2024kd} integrate DP into a CF generator and observe that private CFs reduce KD-based extraction toward no-CF baselines while retaining some agreement and utility. A\"{\i}vodji et al.~\cite{aivodji2020extraction} show that more realistic and diverse CFs improve extraction. Defenders may therefore need to bound the amount, diversity, or precision of recourse information, but doing so can directly conflict with actionability.

\subsection{Prediction-Output Defenses}
Liu et al.'s~\cite{liu2024tellmemore} evaluation of MemGuard~\cite{memguard} is an instructive negative result. MemGuard perturbs confidence scores to confuse a prediction-based membership classifier, but does not protect the feature-importance maps. Liu's explanation-assisted attacks still obtain AUC 0.872--0.885 for LIME, SHAP, Grad-CAM, and SmoothGrad under the evaluated defense. This demonstrates a general principle: protecting one output channel does not protect a parallel explanation channel.

The converse is also true. Explanation sanitization does not necessarily stop prediction-only MIA or extraction. A service that releases both $o(x)$ and $e(x)$ must evaluate the composition. Where multiple explanation formats are available, the strongest joint attack matters, not the strongest single-channel defense.

\subsection{Operational Defenses}
Query budgets, anomaly detection, and extraction defenses such as PRADA~\cite{prada} operate at the service boundary. They are naturally relevant to P1 attacks that require repeated adaptive queries. AUTOLYCUS~\cite{oksuz2024autolycus} observes that budgets can slow extraction. A\"{\i}vodji et al.~\cite{aivodji2020extraction}, however, point out that low-query counterfactual extraction can resemble legitimate use and that a Sybil adversary can distribute queries across identities. As explanations make each request more informative, security cannot rely on volume alone.

P3 secondary disclosure requires a different control set: authentication to stored reports, least-privilege authorization, encryption in transit and at rest, retention limits, and audit logging. These are standard systems controls rather than new XAI algorithms, but the threat is XAI-specific because the explanation report may reveal a private input that the prediction does not. The 2025 scoping review likewise recommends role-based explanation access, query restrictions, and monitoring~\cite{allana2025scoping}; our contribution is to connect each control to the acquisition paths for which it can be effective.

\subsection{Defense Coverage by Acquisition Path}
No single mechanism covers all five acquisition paths. P1 requires explanation minimization or sanitization plus identity-aware query controls; P2 additionally requires defenses on predictions and resistance to surrogate construction; P3 requires treating explanation reports as sensitive data with storage, access, retention, and sharing controls; P4 requires restricting training and model artifacts and auditor privileges; and P5 requires privacy evaluation before irreversible artifact release. Defenses should therefore be composed around the acquisition paths that are present in the deployment.

A defense claim should name the protected asset and the acquisition path it closes. ``Privacy-preserving XAI'' is too broad when a mechanism protects training membership but not query inversion, or sanitizes target explanations while attacker-derived explanations remain available.

\section{Research Gaps and Open Problems}
\label{sec:open}

\paragraph{Standardized marginal and compositional leakage.} The field needs matched experiments in which the target, auxiliary knowledge, attack algorithm, and query budget are fixed while only the explanation channel changes, making $\Delta_{\mathrm{XAI}}$ in Eq.~\ref{eq:delta} measurable. Real services also expose multiple channels; evaluation should test $o(x)$, each $e_i(x)$, and their composition rather than infer safety from one protected output.

\paragraph{Query-record privacy.} Training privacy dominates ML privacy research, but Luo et al.~\cite{luo2022feature}, Zhao et al.~\cite{zhao2021exploiting}, Toma and Kikuchi~\cite{toma2024record}, and Duddu and Boutet~\cite{duddu2022attributes} show that an explanation can leak a record that was never in training. Formal mechanisms need an explicit adjacency or sensitivity notion for the \emph{explained input}, and for any private reference data used to construct the explanation, not only for the model's training set.

\paragraph{Modalities and adaptive acquisition.} Graph extraction has begun to appear~\cite{ma2025egsteal}, but audio, sequential data, and text or LLM explanation interfaces are sparsely covered by mature privacy attacks. At the same time, output sanitization is often evaluated against a fixed attacker; repeated noisy explanations can be aggregated and attackers can switch to P2 surrogate-derived explanations. Both gaps require modality-specific, adaptive evaluation rather than extrapolation from image and tabular results.

\paragraph{Interface-aware leakage benchmarks.} The next step beyond per-paper query counts is a benchmark that fixes the target task and exposes controlled interfaces, for example label only, scores, feature-importance explanations, and counterfactuals, under matched auxiliary knowledge. Such a benchmark would make it possible to compare both marginal leakage and operational cost without pretending that different explanation responses carry equivalent information.

\section{Related Surveys and SoKs}
\label{sec:related}

Nguyen et al.~\cite{nguyen2025survey} survey privacy attacks and countermeasures by explanation type, while Allana et al.~\cite{allana2025scoping} review privacy-preserving XAI more broadly. Nadeem et al.~\cite{nadeem2023sok}, Noppel and Wressnegger~\cite{noppel2024sok}, and Pawlicki et al.~\cite{pawlicki2026weaponising} cover broader XAI security or adversarial themes; Rigaki and Garcia~\cite{rigaki2023survey} survey ML privacy more generally. Our narrower contribution is operational: we separate model knowledge from \emph{how the explanation is acquired}, trace each signal to its attack mechanism and protected asset, and map defenses to the paths they can close.

\section{Conclusion}
\label{sec:conclusion}

The central lesson of this SoK is that the privacy cost of explainability is determined by the \emph{information path}, not by an explanation method alone. Across 25 core studies, explanations reduce extraction cost, expose membership signals, and support reconstruction of private inputs and attributes, but no explanation family is uniformly unsafe. Security claims should identify the protected asset, prediction output, explanation artifact, acquisition path, auxiliary knowledge, and query budget. Defenses require the same discipline: DP training protects training records; sanitization acts on released explanations; service controls constrain online acquisition; and report handling addresses secondary disclosure. This end-to-end view makes attack assumptions, marginal leakage, and defense coverage explicit.

\appendix

\section{Open Science}
This paper systematizes published literature and does not introduce a new attack implementation, dataset, trained model, or benchmark. The underlying evidence is contained in the cited papers and in the study classifications reported in Tables~\ref{tab:xai-signals}--\ref{tab:defenses}. We therefore do not provide a separate research artifact.

\section{Ethical Considerations}
The paper analyzes previously published attacks and defenses and does not introduce a new exploit. Its main risk is that consolidating effective leakage channels could reduce the effort needed to compare prior offensive techniques. We mitigate this by reporting only information already present in the literature and by emphasizing threat-model limits, defensive coverage, and deployment safeguards. The intended benefit is to help model owners and privacy researchers identify which explanation disclosures require protection and avoid overclaiming defenses whose guarantees do not match the protected asset.

\balance
\bibliographystyle{plain}
\bibliography{bibliography_cycle1_13page_final}

@inproceedings{fredrikson2015model,
author = {Fredrikson, Matt and Jha, Somesh and Ristenpart, Thomas},
title = {Model Inversion Attacks That Exploit Confidence Information and Basic Countermeasures},
year = {2015},
isbn = {9781450338325},
publisher = {Association for Computing Machinery},
address = {New York, NY, USA},
url = {https://doi.org/10.1145/2810103.2813677},
doi = {10.1145/2810103.2813677},
booktitle = {Proceedings of the 22nd ACM SIGSAC Conference on Computer and Communications Security},
pages = {1322–1333},
numpages = {12},
location = {Denver, Colorado, USA},
series = {CCS '15}
}

@inproceedings{sundararajan2017axiomatic,
  author       = {Mukund Sundararajan and
                  Ankur Taly and
                  Qiqi Yan},
  editor       = {Doina Precup and
                  Yee Whye Teh},
  title        = {Axiomatic Attribution for Deep Networks},
  booktitle    = {Proceedings of the 34th International Conference on Machine Learning,
                  {ICML} 2017, Sydney, NSW, Australia, 6-11 August 2017},
  series       = {Proceedings of Machine Learning Research},
  volume       = {70},
  pages        = {3319--3328},
  publisher    = {{PMLR}},
  year         = {2017},
  address      = {USA},
  url          = {http://proceedings.mlr.press/v70/sundararajan17a.html},
  bibsource    = {dblp computer science bibliography, https://dblp.org}
}

@article{rigaki2023survey,
  author       = {Maria Rigaki and
                  Sebastian Garc{\'{\i}}a},
  title        = {A Survey of Privacy Attacks in Machine Learning},
  journal      = {{ACM} Comput. Surv.},
  volume       = {56},
  number       = {4},
  pages        = {101:1--101:34},
  year         = {2024},
  url          = {https://doi.org/10.1145/3624010},
  doi          = {10.1145/3624010},
  bibsource    = {dblp computer science bibliography, https://dblp.org}
}

@inproceedings{selvaraju2017gradcam,
  author       = {Ramprasaath R. Selvaraju and
                  Michael Cogswell and
                  Abhishek Das and
                  Ramakrishna Vedantam and
                  Devi Parikh and
                  Dhruv Batra},
  title        = {{Grad-CAM}: Visual Explanations from Deep Networks via Gradient-Based
                  Localization},
  booktitle    = {{IEEE} International Conference on Computer Vision, {ICCV} 2017, Venice,
                  Italy, October 22-29, 2017},
  pages        = {618--626},
  publisher    = {{IEEE} Computer Society},
  address = {USA},
  year         = {2017},
  url          = {https://doi.org/10.1109/ICCV.2017.74},
  doi          = {10.1109/ICCV.2017.74},
  bibsource    = {dblp computer science bibliography, https://dblp.org}
}

@inproceedings{wang2022dualcf,
  author       = {Yongjie Wang and
                  Hangwei Qian and
                  Chunyan Miao},
  title        = {{DualCF}: Efficient Model Extraction Attack from Counterfactual Explanations},
  booktitle    = {FAccT '22: 2022 {ACM} Conference on Fairness, Accountability, and
                  Transparency, Seoul, Republic of Korea, June 21 - 24, 2022},
  pages        = {1318--1329},
  publisher    = {{ACM}},
address = {USA},
  year         = {2022},
  url          = {https://doi.org/10.1145/3531146.3533188},
  doi          = {10.1145/3531146.3533188},
  bibsource    = {dblp computer science bibliography, https://dblp.org}
}

@inproceedings{ribeiro2016lime,
author = {Ribeiro, Marco Tulio and Singh, Sameer and Guestrin, Carlos},
title = {"Why Should I Trust You?": Explaining the Predictions of Any Classifier},
year = {2016},
isbn = {9781450342322},
publisher = {Association for Computing Machinery},
address = {New York, NY, USA},
url = {https://doi.org/10.1145/2939672.2939778},
doi = {10.1145/2939672.2939778},
booktitle = {Proceedings of the 22nd ACM SIGKDD International Conference on Knowledge Discovery and Data Mining},
pages = {1135–1144},
numpages = {10},
location = {San Francisco, California, USA},
series = {KDD '16}
}

@misc{aivodji2020extraction,
  author    = {Ulrich A{\"{\i}}vodji and
               Alexandre Bolot and
               S{\'{e}}bastien Gambs},
  title     = {Model extraction from counterfactual explanations},
  journal   = {CoRR},
  volume    = {abs/2009.01884},
  year      = {2020},
  url       = {https://doi.org/10.48550/arXiv.2009.01884},
  eprinttype = {arXiv},
  eprint    = {2009.01884},
  bibsource = {dblp computer science bibliography, https://dblp.org}
}

@article{bach2015lrp,
  title={On pixel-wise explanations for non-linear classifier decisions by layer-wise relevance propagation},
  author={Bach, Sebastian and Binder, Alexander and Montavon, Gr{\'e}goire and Klauschen, Frederick and M{\"u}ller, Klaus-Robert and Samek, Wojciech},
  journal={PloS one},
  volume={10},
  number={7},
  pages={e0130140},
  year={2015},
  publisher={Public Library of Science San Francisco, CA USA},
  url = {https://doi.org/10.1371/journal.pone.0130140}
}

@inproceedings{lundberg2017shap,
author = {Lundberg, Scott M. and Lee, Su-In},
title = {A Unified Approach to Interpreting Model Predictions},
year = {2017},
isbn = {9781510860964},
publisher = {Curran Associates Inc.},
address = {Red Hook, NY, USA},
booktitle = {Proceedings of the 31st International Conference on Neural Information Processing Systems},
pages = {4768–4777},
numpages = {10},
location = {Long Beach, California, USA},
series = {NIPS'17},
url = {https://dl.acm.org/doi/pdf/10.5555/3295222.3295230}
}

@inproceedings{milli2019reconstruction,
author = {Milli, Smitha and Schmidt, Ludwig and Dragan, Anca D. and Hardt, Moritz},
title = {Model Reconstruction from Model Explanations},
year = {2019},
isbn = {9781450361255},
publisher = {Association for Computing Machinery},
address = {New York, NY, USA},
url = {https://doi.org/10.1145/3287560.3287562},
doi = {10.1145/3287560.3287562},
booktitle = {Proceedings of the Conference on Fairness, Accountability, and Transparency},
pages = {1–9},
numpages = {9},
location = {Atlanta, GA, USA},
series = {FAT* '19}
}

@INPROCEEDINGS {liu2024tellmemore,
author = {Liu, Han and Wu, Yuhao and Yu, Zhiyuan and Zhang, Ning },
booktitle = { 2024 IEEE Symposium on Security and Privacy (SP) },
title = {{ Please Tell Me More: Privacy Impact of Explainability through the Lens of Membership Inference Attack }},
year = {2024},
volume = {},
ISSN = {},
pages = {4791-4809},
doi = {10.1109/SP54263.2024.00120},
url = {https://doi.ieeecomputersociety.org/10.1109/SP54263.2024.00120},
publisher = {IEEE Computer Society},
address = {Los Alamitos, CA, USA},
month =May}

@inproceedings{dpsgd,
author = {Abadi, Martin and Chu, Andy and Goodfellow, Ian and McMahan, H. Brendan and Mironov, Ilya and Talwar, Kunal and Zhang, Li},
title = {Deep Learning with Differential Privacy},
year = {2016},
isbn = {9781450341394},
publisher = {Association for Computing Machinery},
address = {New York, NY, USA},
url = {https://doi.org/10.1145/2976749.2978318},
doi = {10.1145/2976749.2978318},
booktitle = {Proceedings of the 2016 ACM SIGSAC Conference on Computer and Communications Security},
pages = {308–318},
numpages = {11},
location = {Vienna, Austria},
series = {CCS '16}
}

@inproceedings{prada,
  author       = {Mika Juuti and
                  Sebastian Szyller and
                  Samuel Marchal and
                  N. Asokan},
  title        = {{PRADA:} Protecting Against {DNN} Model Stealing Attacks},
  booktitle    = {{IEEE} European Symposium on Security and Privacy, EuroS{\&}P
                  2019, Stockholm, Sweden, June 17-19, 2019},
  pages        = {512--527},
  publisher    = {{IEEE}},
  address = {USA},
  year         = {2019},
  url          = {https://doi.org/10.1109/EuroSP.2019.00044},
  doi          = {10.1109/EUROSP.2019.00044},
  bibsource    = {dblp computer science bibliography, https://dblp.org}
}

@article{dropout,
  author       = {Nitish Srivastava and
                  Geoffrey E. Hinton and
                  Alex Krizhevsky and
                  Ilya Sutskever and
                  Ruslan Salakhutdinov},
  title        = {Dropout: a simple way to prevent neural networks from overfitting},
  journal      = {J. Mach. Learn. Res.},
  volume       = {15},
  number       = {1},
  pages        = {1929--1958},
  year         = {2014},
  url          = {https://dl.acm.org/doi/10.5555/2627435.2670313},
  doi          = {10.5555/2627435.2670313},
  bibsource    = {dblp computer science bibliography, https://dblp.org}
}

@inproceedings{memguard,
author = {Jia, Jinyuan and Salem, Ahmed and Backes, Michael and Zhang, Yang and Gong, Neil Zhenqiang},
title = {{MemGuard}: Defending against Black-Box Membership Inference Attacks via Adversarial Examples},
year = {2019},
isbn = {9781450367479},
publisher = {Association for Computing Machinery},
address = {New York, NY, USA},
url = {https://doi.org/10.1145/3319535.3363201},
doi = {10.1145/3319535.3363201},
booktitle = {Proceedings of the 2019 ACM SIGSAC Conference on Computer and Communications Security},
pages = {259–274},
numpages = {16},
location = {London, United Kingdom},
series = {CCS '19}
}

@article{oksuz2024autolycus,
  author       = {Abdullah Caglar Oksuz and
                  Anisa Halimi and
                  Erman Ayday},
  title        = {{AUTOLYCUS:} Exploiting Explainable Artificial Intelligence {(XAI)} for Model Extraction Attacks against Interpretable Models},
  journal      = {Proc. Priv. Enhancing Technol.},
  volume       = {2024},
  number       = {4},
  pages        = {684--699},
  year         = {2024},
  url          = {https://doi.org/10.56553/popets-2024-0137},
  doi          = {10.56553/POPETS-2024-0137},
  bibsource    = {dblp computer science bibliography, https://dblp.org}
}

@inproceedings{shokri2021privacy,
  author       = {Reza Shokri and
                  Martin Strobel and
                  Yair Zick},
  editor       = {Marion Fourcade and
                  Benjamin Kuipers and
                  Seth Lazar and
                  Deirdre K. Mulligan},
  title        = {On the Privacy Risks of Model Explanations},
  booktitle    = {{AIES} '21: {AA{AI}/ACM} Conference on {AI}, Ethics, and Society, Virtual
                  Event, USA, May 19-21, 2021},
  pages        = {231--241},
  publisher    = {{ACM}},
address = {USA},
  year         = {2021},
  url          = {https://doi.org/10.1145/3461702.3462533},
  doi          = {10.1145/3461702.3462533},
  bibsource    = {dblp computer science bibliography, https://dblp.org}
}

@inproceedings{zhao2021exploiting,
  author       = {Xuejun Zhao and
                  Wencan Zhang and
                  Xiaokui Xiao and
                  Brian Y. Lim},
  title        = {Exploiting Explanations for Model Inversion Attacks},
  booktitle    = {2021 {IEEE/CVF} International Conference on Computer Vision, {ICCV}
                  2021, Montreal, QC, Canada, October 10-17, 2021},
  pages        = {662--672},
  publisher    = {{IEEE}},
  address = {USA},
  year         = {2021},
  url          = {https://doi.org/10.1109/ICCV48922.2021.00072},
  doi          = {10.1109/ICCV48922.2021.00072},
  bibsource    = {dblp computer science bibliography, https://dblp.org}
}

@article{yan2023xamea,
title = {Explanation leaks: Explanation-guided model extraction attacks},
journal = {Information Sciences},
volume = {632},
pages = {269-284},
year = {2023},
issn = {0020-0255},
doi = {https://doi.org/10.1016/j.ins.2023.03.020},
url = {https://www.sciencedirect.com/science/article/pii/S002002552300316X},
author = {Anli Yan and Teng Huang and Lishan Ke and Xiaozhang Liu and Qi Chen and Changyu Dong}
}

@inproceedings{miura2021megex,
author = {Miura, Takayuki and Shibahara, Toshiki and Yanai, Naoto},
title = {{MEGEX}: Data-Free Model Extraction Attack Against Gradient-Based Explainable {AI}},
year = {2024},
isbn = {9798400706912},
publisher = {Association for Computing Machinery},
address = {New York, NY, USA},
url = {https://doi.org/10.1145/3665451.3665533},
doi = {10.1145/3665451.3665533},
booktitle = {Proceedings of the 2nd ACM Workshop on Secure and Trustworthy Deep Learning Systems},
pages = {56–66},
numpages = {11},
location = {Singapore, Singapore},
series = {SecTL '24}
}

@article{nguyen2025survey,
  author       = {Thanh Tam Nguyen and
                  Thanh Trung Huynh and
                  Zhao Ren and
                  Thanh Toan Nguyen and
                  Phi Le Nguyen and
                  Hongzhi Yin and
                  Quoc Viet Hung Nguyen},
  title        = {Privacy-preserving explainable {AI:} a survey},
  journal      = {Sci. China Inf. Sci.},
  volume       = {68},
  number       = {1},
  year         = {2025},
  url          = {https://doi.org/10.1007/s11432-024-4123-4},
  doi          = {10.1007/S11432-024-4123-4},
  bibsource    = {dblp computer science bibliography, https://dblp.org}
}

@inproceedings{allana2025noise,
  author       = {Sonal Allana and
                  Rozita Dara},
  title        = {Privacy Preservation with Noise in Explainable {AI}},
  booktitle    = {22nd Annual International Conference on Privacy, Security, and Trust,
                  {PST} 2025, Fredericton, NB, Canada, August 26-28, 2025},
  pages        = {1--8},
  publisher    = {{IEEE}},
  year         = {2025},
  url          = {https://doi.org/10.1109/PST65910.2025.11268815},
  doi          = {10.1109/PST65910.2025.11268815},
  bibsource    = {dblp computer science bibliography, https://dblp.org}
}

@article{wachter2018counterfactual,
  author    = {Wachter, Sandra and Mittelstadt, Brent and Russell, Chris},
  title     = {Counterfactual Explanations without Opening the Black Box: Automated Decisions and the {G{DP}R}},
  journal   = {Harvard Journal of Law \& Technology},
  volume    = {31},
  number    = {2},
  pages     = {841--887},
  year      = {2018},
  url       = {https://doi.org/10.2139/ssrn.3063289}
}

@inproceedings{mothilal2020dice,
  author       = {Ramaravind Kommiya Mothilal and
                  Amit Sharma and
                  Chenhao Tan},
  editor       = {Mireille Hildebrandt and
                  Carlos Castillo and
                  L. Elisa Celis and
                  Salvatore Ruggieri and
                  Linnet Taylor and
                  Gabriela Zanfir{-}Fortuna},
  title        = {Explaining machine learning classifiers through diverse counterfactual
                  explanations},
  booktitle    = {FAT* '20: Conference on Fairness, Accountability, and Transparency,
                  Barcelona, Spain, January 27-30, 2020},
  pages        = {607--617},
  publisher    = {{ACM}},
  year         = {2020},
  url          = {https://doi.org/10.1145/3351095.3372850},
  doi          = {10.1145/3351095.3372850},
  bibsource    = {dblp computer science bibliography, https://dblp.org}
}

@inproceedings{koh2017influence,
  author       = {Pang Wei Koh and
                  Percy Liang},
  editor       = {Doina Precup and
                  Yee Whye Teh},
  title        = {Understanding Black-box Predictions via Influence Functions},
  booktitle    = {Proceedings of the 34th International Conference on Machine Learning,
                  {ICML} 2017, Sydney, NSW, Australia, 6-11 August 2017},
  series       = {Proceedings of Machine Learning Research},
  volume       = {70},
  pages        = {1885--1894},
  publisher    = {{PMLR}},
  year         = {2017},
  url          = {http://proceedings.mlr.press/v70/koh17a.html},
  bibsource    = {dblp computer science bibliography, https://dblp.org}
}

@inproceedings{ezzeddine2024kd,
    title={Knowledge Distillation-Based Model Extraction Attack using {GAN}-based Private Counterfactual Explanations},
    author={Fatima Ezzeddine and Omran Ayoub and Silvia Giordano},
    booktitle={NeurIPS 2024 Workshop on Regulatable ML},
    year={2024},
    url={https://openreview.net/forum?id=bGtv9Lcw80}
}

@article{ezzeddine2026leak,
  author       = {Fatima Ezzeddine and
                  Osama Zammar and
                  Silvia Giordano and
                  Omran Ayoub},
  title        = {Explanations Leak: Membership Inference with Differential Privacy
                  and Active Learning Defense},
  journal      = {CoRR},
  volume       = {abs/2602.03611},
  year         = {2026},
  url          = {https://doi.org/10.48550/arXiv.2602.03611},
  doi          = {10.48550/ARXIV.2602.03611},
  eprinttype   = {arXiv},
  eprint       = {2602.03611},
  bibsource    = {dblp computer science bibliography, https://dblp.org}
}

@inproceedings{luo2022feature,
  author       = {Xinjian Luo and
                  Yangfan Jiang and
                  Xiaokui Xiao},
  editor       = {Heng Yin and
                  Angelos Stavrou and
                  Cas Cremers and
                  Elaine Shi},
  title        = {Feature Inference Attack on Shapley Values},
  booktitle    = {Proceedings of the 2022 {ACM} {SIGSAC} Conference on Computer and
                  Communications Security, {CCS} 2022, Los Angeles, CA, USA, November
                  7-11, 2022},
  pages        = {2233--2247},
  publisher    = {{ACM}},
  year         = {2022},
  url          = {https://doi.org/10.1145/3548606.3560573},
  doi          = {10.1145/3548606.3560573},
  bibsource    = {dblp computer science bibliography, https://dblp.org}
}

@inproceedings{toma2024record,
  author       = {Ryotaro Toma and
                  Hiroaki Kikuchi},
  editor       = {Josep Domingo{-}Ferrer and
                  Melek {\"{O}}nen},
  title        = {Combinations of {AI} Models and {XAI} Metrics Vulnerable to Record
                  Reconstruction Risk},
  booktitle    = {Privacy in Statistical Databases - International Conference, {PSD}
                  2024, Antibes Juan-les-Pins, France, September 25-27, 2024, Proceedings},
  series       = {Lecture Notes in Computer Science},
  volume       = {14915},
  pages        = {329--343},
  publisher    = {Springer},
  year         = {2024},
  url          = {https://doi.org/10.1007/978-3-031-69651-0\_22},
  doi          = {10.1007/978-3-031-69651-0\_22},
  bibsource    = {dblp computer science bibliography, https://dblp.org}
}

@inproceedings{cohen2024influence,
  author       = {Gilad Cohen and
                  Raja Giryes},
  title        = {Membership Inference Attack Using Self Influence Functions},
  booktitle    = {{IEEE/CVF} Winter Conference on Applications of Computer Vision, {WACV}
                  2024, Waikoloa, HI, USA, January 3-8, 2024},
  pages        = {4880--4889},
  publisher    = {{IEEE}},
  year         = {2024},
  url          = {https://doi.org/10.1109/WACV57701.2024.00482},
  doi          = {10.1109/WACV57701.2024.00482},
  bibsource    = {dblp computer science bibliography, https://dblp.org}
}

@inproceedings{khouna2025trees,
 author = {Khouna, Awa and Ferry, Julien and Vidal, Thibaut},
 booktitle = {Advances in Neural Information Processing Systems},
 editor = {D. Belgrave and C. Zhang and H. Lin and R. Pascanu and P. Koniusz and M. Ghassemi and N. Chen},
 pages = {2721--2756},
 publisher = {Curran Associates, Inc.},
 title = {From Counterfactuals to Trees: Competitive Analysis of Model Extraction Attacks},
 url = {https://proceedings.neurips.cc/paper_files/paper/2025/file/040d3b6af368bf71f952c18da5713b48-Paper-Conference.pdf},
 volume = {38},
 year = {2025}
}

@inproceedings{pawelczyk2023recourse,
  author       = {Martin Pawelczyk and
                  Himabindu Lakkaraju and
                  Seth Neel},
  editor       = {Francisco J. R. Ruiz and
                  Jennifer G. Dy and
                  Jan{-}Willem van de Meent},
  title        = {On the Privacy Risks of Algorithmic Recourse},
  booktitle    = {International Conference on Artificial Intelligence and Statistics,
                  25-27 April 2023, Palau de Congressos, Valencia, Spain},
  series       = {Proceedings of Machine Learning Research},
  volume       = {206},
  pages        = {9680--9696},
  publisher    = {{PMLR}},
  year         = {2023},
  url          = {https://proceedings.mlr.press/v206/pawelczyk23a.html},
  bibsource    = {dblp computer science bibliography, https://dblp.org}
}

@article{otto2026linear,
  author       = {Daan Otto and
                  Jannis Kurtz and
                  Dick den Hertog and
                  S. Ilker Birbil},
  title        = {Linear Model Extraction via Factual and Counterfactual Queries},
  journal      = {CoRR},
  volume       = {abs/2602.09748},
  year         = {2026},
  url          = {https://doi.org/10.48550/arXiv.2602.09748},
  doi          = {10.48550/ARXIV.2602.09748},
  eprinttype   = {arXiv},
  eprint       = {2602.09748},
  bibsource    = {dblp computer science bibliography, https://dblp.org}
}

@article{ma2024labelonly,
  author       = {Yao Ma and
                  Xurong Zhai and
                  Dan Yu and
                  Yuli Yang and
                  Xingyu Wei and
                  Yongle Chen},
  title        = {Label-Only Membership Inference Attack Based on Model Explanation},
  journal      = {Neural Process. Lett.},
  volume       = {56},
  number       = {5},
  pages        = {236},
  year         = {2024},
  url          = {https://doi.org/10.1007/s11063-024-11682-1},
  doi          = {10.1007/S11063-024-11682-1},
  bibsource    = {dblp computer science bibliography, https://dblp.org}
}

@article{ma2025egsteal,
  author       = {Bin Ma and
                  Yuyuan Feng and
                  Minhua Lin and
                  Enyan Dai},
  title        = {How Explanations Leak the Decision Logic: Stealing Graph Neural Networks
                  via Explanation Alignment},
  journal      = {CoRR},
  volume       = {abs/2506.03087},
  year         = {2025},
  url          = {https://doi.org/10.48550/arXiv.2506.03087},
  doi          = {10.48550/ARXIV.2506.03087},
  eprinttype   = {arXiv},
  eprint       = {2506.03087},
  bibsource    = {dblp computer science bibliography, https://dblp.org}
}

@article{yan2023dmeae,
  author       = {Anli Yan and
                  Ruitao Hou and
                  Hongyang Yan and
                  Xiaozhang Liu},
  title        = {Explanation-based data-free model extraction attacks},
  journal      = {World Wide Web {(WWW)}},
  volume       = {26},
  number       = {5},
  pages        = {3081--3092},
  year         = {2023},
  url          = {https://doi.org/10.1007/s11280-023-01150-6},
  doi          = {10.1007/S11280-023-01150-6},
  bibsource    = {dblp computer science bibliography, https://dblp.org}
}

@article{yan2022dtmea,
  author       = {Anli Yan and
                  Ruitao Hou and
                  Xiaozhang Liu and
                  Hongyang Yan and
                  Teng Huang and
                  Xianmin Wang},
  title        = {Towards explainable model extraction attacks},
  journal      = {Int. J. Intell. Syst.},
  volume       = {37},
  number       = {11},
  pages        = {9936--9956},
  year         = {2022},
  url          = {https://doi.org/10.1002/int.23022},
  doi          = {10.1002/INT.23022},
  bibsource    = {dblp computer science bibliography, https://dblp.org}
}

@inproceedings{naretto2022global,
  author       = {Francesca Naretto and
                  Anna Monreale and
                  Fosca Giannotti},
  title        = {Evaluating the Privacy Exposure of Interpretable Global Explainers},
  booktitle    = {4th {IEEE} International Conference on Cognitive Machine Intelligence,
                  CogMI 2022, Atlanta, GA, USA, December 14-17, 2022},
  pages        = {13--19},
  publisher    = {{IEEE}},
  year         = {2022},
  url          = {https://doi.org/10.1109/CogMI56440.2022.00012},
  doi          = {10.1109/COGMI56440.2022.00012},
  bibsource    = {dblp computer science bibliography, https://dblp.org}
}

@inproceedings{wainakh2021llg,
  author       = {Aidmar Wainakh and
                  Till M{\"{u}}{\ss}ig and
                  Tim Grube and
                  Max M{\"{u}}hlh{\"{a}}user},
  title        = {Label Leakage from Gradients in Distributed Machine Learning},
  booktitle    = {18th {IEEE} Annual Consumer Communications {\&} Networking Conference,
                  {CCNC} 2021, Las Vegas, NV, USA, January 9-12, 2021},
  pages        = {1--4},
  publisher    = {{IEEE}},
  year         = {2021},
  url          = {https://doi.org/10.1109/CCNC49032.2021.9369498},
  doi          = {10.1109/CCNC49032.2021.9369498},
  bibsource    = {dblp computer science bibliography, https://dblp.org}
}

@article{zhao2020idlg,
  author       = {Bo Zhao and
                  Konda Reddy Mopuri and
                  Hakan Bilen},
  title        = {{iDLG}: Improved Deep Leakage from Gradients},
  journal      = {CoRR},
  volume       = {abs/2001.02610},
  year         = {2020},
  url          = {http://arxiv.org/abs/2001.02610},
  eprinttype   = {arXiv},
  eprint       = {2001.02610},
  bibsource    = {dblp computer science bibliography, https://dblp.org}
}

@inproceedings{zhu2019deep,
  author       = {Ligeng Zhu and
                  Zhijian Liu and
                  Song Han},
  editor       = {Hanna M. Wallach and
                  Hugo Larochelle and
                  Alina Beygelzimer and
                  Florence d'Alch{\'{e}}{-}Buc and
                  Emily B. Fox and
                  Roman Garnett},
  title        = {Deep Leakage from Gradients},
  booktitle    = {Advances in Neural Information Processing Systems 32: Annual Conference
                  on Neural Information Processing Systems 2019, NeurIPS 2019, December
                  8-14, 2019, Vancouver, BC, Canada},
  pages        = {14747--14756},
  year         = {2019},
  url          = {https://proceedings.neurips.cc/paper/2019/hash/60a6c4002cc7b29142def8871531281a-Abstract.html},
  bibsource    = {dblp computer science bibliography, https://dblp.org}
}

@article{kuppa2021adversarial,
  author       = {Aditya Kuppa and
                  Nhien{-}An Le{-}Khac},
  title        = {Adversarial {XAI} Methods in Cybersecurity},
  journal      = {{IEEE} Trans. Inf. Forensics Secur.},
  volume       = {16},
  pages        = {4924--4938},
  year         = {2021},
  url          = {https://doi.org/10.1109/TIFS.2021.3117075},
  doi          = {10.1109/TIFS.2021.3117075},
  bibsource    = {dblp computer science bibliography, https://dblp.org}
}

@inproceedings{duddu2022attributes,
  author       = {Vasisht Duddu and
                  Antoine Boutet},
  editor       = {Mohammad Al Hasan and
                  Li Xiong},
  title        = {Inferring Sensitive Attributes from Model Explanations},
  booktitle    = {Proceedings of the 31st {ACM} International Conference on Information
                  {\&} Knowledge Management, Atlanta, GA, USA, October 17-21, 2022},
  pages        = {416--425},
  publisher    = {{ACM}},
  year         = {2022},
  url          = {https://doi.org/10.1145/3511808.3557362},
  doi          = {10.1145/3511808.3557362},
  bibsource    = {dblp computer science bibliography, https://dblp.org}
}

@article{allana2025scoping,
  author       = {Sonal Allana and
                  Mohan Kankanhalli and
                  Rozita Dara},
  title        = {Privacy Risks and Preservation Methods in Explainable Artificial Intelligence:
                  {A} Scoping Review},
  journal      = {Trans. Mach. Learn. Res.},
  volume       = {2025},
  year         = {2025},
  url          = {https://openreview.net/forum?id=q9nykJfzku},
  bibsource    = {dblp computer science bibliography, https://dblp.org}
}

@inproceedings{nguyen2023xrand,
  author       = {Truc D. T. Nguyen and
                  Phung Lai and
                  Hai Phan and
                  My T. Thai},
  editor       = {Brian Williams and
                  Yiling Chen and
                  Jennifer Neville},
  title        = {{XRand}: Differentially Private Defense against Explanation-Guided Attacks},
  booktitle    = {Thirty-Seventh {AAAI} Conference on Artificial Intelligence, {AAAI}
                  2023, Thirty-Fifth Conference on Innovative Applications of Artificial
                  Intelligence, {IAAI} 2023, Thirteenth Symposium on Educational Advances
                  in Artificial Intelligence, {EAAI} 2023, Washington, DC, USA, February
                  7-14, 2023},
  pages        = {11873--11881},
  publisher    = {{AAAI} Press},
  year         = {2023},
  url          = {https://doi.org/10.1609/aaai.v37i10.26401},
  doi          = {10.1609/AAAI.V37I10.26401},
  bibsource    = {dblp computer science bibliography, https://dblp.org}
}

@inproceedings{patel2022dp,
  author       = {Neel Patel and
                  Reza Shokri and
                  Yair Zick},
  title        = {Model Explanations with Differential Privacy},
  booktitle    = {FAccT '22: 2022 {ACM} Conference on Fairness, Accountability, and
                  Transparency, Seoul, Republic of Korea, June 21 - 24, 2022},
  pages        = {1895--1904},
  publisher    = {{ACM}},
  year         = {2022},
  url          = {https://doi.org/10.1145/3531146.3533235},
  doi          = {10.1145/3531146.3533235},
  bibsource    = {dblp computer science bibliography, https://dblp.org}
}

@inproceedings{zhong2025faithful,
  author       = {Chudi Zhong and
                  Panyu Chen and
                  Cynthia Rudin},
  editor       = {Yingzhen Li and
                  Stephan Mandt and
                  Shipra Agrawal and
                  Mohammad Emtiyaz Khan},
  title        = {Models That Are Interpretable But Not Transparent},
  booktitle    = {International Conference on Artificial Intelligence and Statistics,
                  {AISTATS} 2025, Mai Khao, Thailand, 3-5 May 2025},
  series       = {Proceedings of Machine Learning Research},
  volume       = {258},
  pages        = {1648--1656},
  publisher    = {{PMLR}},
  year         = {2025},
  url          = {https://proceedings.mlr.press/v258/zhong25b.html},
  bibsource    = {dblp computer science bibliography, https://dblp.org}
}

@inproceedings{pawlicki2026weaponising,
  author       = {Marek Pawlicki and
                  Ryszard S. Choras and
                  Rafal Kozik and
                  Michal Choras},
  editor       = {Ana Paula Rocha and
                  Mattias Wahde and
                  H. Jaap van den Herik},
  title        = {Survey on Explainability-Weaponising Adversarial Attack Vectors against
                  Deep Neural Networks and Artificial Intelligence},
  booktitle    = {Proceedings of the 18th International Conference on Agents and Artificial
                  Intelligence, {ICAART} 2026 - Volume 2, Marbella, Spain, March 5-7,
                  2026},
  pages        = {1361--1370},
  publisher    = {{SCITEPRESS}},
  year         = {2026},
  url          = {https://doi.org/10.5220/0014286400004052},
  doi          = {10.5220/0014286400004052},
  bibsource    = {dblp computer science bibliography, https://dblp.org}
}

@article{goethals2023linkage,
  author       = {Sofie Goethals and
                  Kenneth S{\"{o}}rensen and
                  David Martens},
  title        = {The Privacy Issue of Counterfactual Explanations: Explanation Linkage
                  Attacks},
  journal      = {{ACM} Trans. Intell. Syst. Technol.},
  volume       = {14},
  number       = {5},
  pages        = {83:1--83:24},
  year         = {2023},
  url          = {https://doi.org/10.1145/3608482},
  doi          = {10.1145/3608482},
  bibsource    = {dblp computer science bibliography, https://dblp.org}
}

@inproceedings{nadeem2023sok,
  author    = {Azqa Nadeem and Dani{\"e}l Vos and Clinton Cao and Luca Pajola and Simon Dieck and Robert Baumgartner and Sicco Verwer},
  title     = {{SoK: Explainable Machine Learning for Computer Security Applications}},
  booktitle = {2023 IEEE 8th European Symposium on Security and Privacy (EuroS\&P)},
  pages     = {221--240},
  year      = {2023},
  doi       = {10.1109/EuroSP57164.2023.00022}
}

@inproceedings{noppel2024sok,
  author    = {Maximilian Noppel and Christian Wressnegger},
  title     = {{SoK: Explainable Machine Learning in Adversarial Environments}},
  booktitle = {2024 IEEE Symposium on Security and Privacy (SP)},
  pages     = {2441--2459},
  year      = {2024},
  doi       = {10.1109/SP54263.2024.00021}
}

@inproceedings{benhmida2026deepleak,
  author    = {Firas Ben Hmida and Zain Sbeih and Philemon Hailemariam and Birhanu Eshete},
  title     = {{DeepLeak: Privacy Enhancing Hardening of Model Explanations Against Membership Leakage}},
  booktitle = {2026 IEEE Conference on Secure and Trustworthy Machine Learning (SaTML)},
  year      = {2026},
  url       = {https://arxiv.org/abs/2601.03429}
}
\end{document}